\documentclass[longauth]{aa}
\usepackage{txfonts,graphicx,wasysym}

\usepackage{siunitx}
\usepackage{booktabs}
\usepackage{hyperref}
\hypersetup{colorlinks=true,linkcolor=blue,filecolor=blue,urlcolor=blue,citecolor=blue}
\usepackage{xcolor}
\usepackage{comment}
\bibpunct{(}{)}{;}{a}{}{,}
\usepackage{amssymb}
\usepackage{amsmath}
\usepackage{longtable}
\usepackage{appendix}
\usepackage{lscape}
\usepackage[switch]{lineno} 

\usepackage{txfonts}
\makeatletter
\renewcommand*\aa@pageof{, page \thepage{} of \pageref*{LastPage}}
\makeatother

\begin{document}
\title{Testing the ubiquitous presence of very high energy emission in gamma-ray bursts with the MAGIC telescopes}
\titlerunning{MAGIC observations of gamma-ray bursts}

%
\author{ S.~Abe\inst{1} \and
J.~Abhir\inst{2} \and
A.~Abhishek\inst{3} \and
V.~A.~Acciari\inst{4} \and
A.~Aguasca-Cabot\inst{5} \and
I.~Agudo\inst{6} \and
T.~Aniello\inst{7} \and
S.~Ansoldi\inst{8,43} \and
L.~A.~Antonelli\inst{7} \and
A.~Arbet Engels\inst{9}
% ...
}
\institute { Japanese MAGIC Group: Institute for Cosmic Ray Research (ICRR), The University of Tokyo, Kashiwa, 277-8582 Chiba, Japan
\and ETH Z\"urich, CH-8093 Z\"urich, Switzerland
\and Universit\`a di Siena and INFN Pisa, I-53100 Siena, Italy
\and Institut de F\'isica d'Altes Energies (IFAE), The Barcelona Institute of Science and Technology (BIST), E-08193 Bellaterra (Barcelona), Spain
\and Universitat de Barcelona, ICCUB, IEEC-UB, E-08028 Barcelona, Spain
\and Instituto de Astrof\'isica de Andaluc\'ia-CSIC, Glorieta de la Astronom\'ia s/n, 18008, Granada, Spain
\and National Institute for Astrophysics (INAF), I-00136 Rome, Italy
\and Universit\`a di Udine and INFN Trieste, I-33100 Udine, Italy
\and Max-Planck-Institut f\"ur Physik, D-85748 Garching, Germany
\and Universit\`a di Padova and INFN, I-35131 Padova, Italy
\and Instituto de Astrof\'isica de Canarias and Dpto. de  Astrof\'isica, Universidad de La Laguna, E-38200, La Laguna, Tenerife, Spain
\and Croatian MAGIC Group: University of Zagreb, Faculty of Electrical Engineering and Computing (FER), 10000 Zagreb, Croatia
\and Saha Institute of Nuclear Physics, A CI of Homi Bhabha National Institute, Kolkata 700064, West Bengal, India
\and Centro Brasileiro de Pesquisas F\'isicas (CBPF), 22290-180 URCA, Rio de Janeiro (RJ), Brazil
\and IPARCOS Institute and EMFTEL Department, Universidad Complutense de Madrid, E-28040 Madrid, Spain
\and University of Lodz, Faculty of Physics and Applied Informatics, Department of Astrophysics, 90-236 Lodz, Poland
\and Centro de Investigaciones Energ\'eticas, Medioambientales y Tecnol\'ogicas, E-28040 Madrid, Spain
\and Departament de F\'isica, and CERES-IEEC, Universitat Aut\`onoma de Barcelona, E-08193 Bellaterra, Spain
\and Universit\`a di Pisa and INFN Pisa, I-56126 Pisa, Italy
\and INFN MAGIC Group: INFN Sezione di Bari and Dipartimento Interateneo di Fisica dell'Universit\`a e del Politecnico di Bari, I-70125 Bari, Italy
\and Department for Physics and Technology, University of Bergen, Norway
\and INFN MAGIC Group: INFN Sezione di Torino and Universit\`a degli Studi di Torino, I-10125 Torino, Italy
\and Croatian MAGIC Group: University of Rijeka, Faculty of Physics, 51000 Rijeka, Croatia
\and Universit\"at W\"urzburg, D-97074 W\"urzburg, Germany
\and Technische Universit\"at Dortmund, D-44221 Dortmund, Germany
\and Japanese MAGIC Group: Physics Program, Graduate School of Advanced Science and Engineering, Hiroshima University, 739-8526 Hiroshima, Japan
\and Deutsches Elektronen-Synchrotron (DESY), D-15738 Zeuthen, Germany
\and Armenian MAGIC Group: ICRANet-Armenia, 0019 Yerevan, Armenia
\and Croatian MAGIC Group: University of Split, Faculty of Electrical Engineering, Mechanical Engineering and Naval Architecture (FESB), 21000 Split, Croatia
\and Croatian MAGIC Group: Josip Juraj Strossmayer University of Osijek, Department of Physics, 31000 Osijek, Croatia
\and Finnish MAGIC Group: Finnish Centre for Astronomy with ESO, Department of Physics and Astronomy, University of Turku, FI-20014 Turku, Finland
\and Japanese MAGIC Group: Department of Physics, Tokai University, Hiratsuka, 259-1292 Kanagawa, Japan
\and University of Geneva, Chemin d'Ecogia 16, CH-1290 Versoix, Switzerland
\and Inst. for Nucl. Research and Nucl. Energy, Bulgarian Academy of Sciences, BG-1784 Sofia, Bulgaria
\and INFN MAGIC Group: INFN Sezione di Catania and Dipartimento di Fisica e Astronomia, University of Catania, I-95123 Catania, Italy
\and Japanese MAGIC Group: Department of Physics, Yamagata University, Yamagata 990-8560, Japan
\and Finnish MAGIC Group: Space Physics and Astronomy Research Unit, University of Oulu, FI-90014 Oulu, Finland
\and Japanese MAGIC Group: Chiba University, ICEHAP, 263-8522 Chiba, Japan
\and Japanese MAGIC Group: Institute for Space-Earth Environmental Research and Kobayashi-Maskawa Institute for the Origin of Particles and the Universe, Nagoya University, 464-6801 Nagoya, Japan
\and Japanese MAGIC Group: Department of Physics, Kyoto University, 606-8502 Kyoto, Japan
\and INFN MAGIC Group: INFN Roma Tor Vergata, I-00133 Roma, Italy
\and Japanese MAGIC Group: Department of Physics, Konan University, Kobe, Hyogo 658-8501, Japan
\and also at International Center for Relativistic Astrophysics (ICRA), Rome, Italy
% ...
}

\date{Received: 9 May 2025~/~Accepted: 27 June 2025 \\}

\abstract 
{Gamma-ray bursts (GRBs) are the most powerful transient objects in the Universe, and they are a primary target for the MAGIC Collaboration. Recognizing the challenges of observing these elusive objects with Imaging Atmospheric Cherenkov Telescopes (IACTs), we implemented a dedicated observational strategy that included an automated procedure for rapid re-pointing to transient sources. Since 2013, this automated procedure has enabled MAGIC to observe GRBs at a rate of approximately ten per year, which led to the successful detection of two GRBs at very high energies (VHE; E > 100 GeV). We present a comprehensive analysis of 42 non-detected GRBs (4 short GRBs) observed by MAGIC from 2013 to 2019. We derived upper limits (ULs) on the observed energy flux as well as on the intrinsic energy flux corrected for absorption by the extragalactic background light (EBL) from the MAGIC observations in selected energy and time intervals. We conducted a comprehensive study of their properties to investigate the reasons for these non-detections, including the possible peculiar properties of TeV-detected GRBs. We find that strong EBL absorption significantly hinders TeV detection for the majority of GRBs in our sample. For a subset of 6 GRBs with redshift $z<2$, we compared the UL on the intrinsic flux in the VHE domain with the simultaneous X-ray flux, which is observed to be at the same level in the current population of TeV-detected GRBs. Based on these inferred MAGIC ULs, we conclude that a VHE component with a luminosity comparable to the simultaneously observed X-ray luminosity cannot be ruled out for this sample.}


\authorrunning{Abe et al.}

\maketitle

\section{Introduction}
\label{sec:intro}
Gamma-ray bursts (GRBs) represent one of the most enigmatic classes of transient sources~\citep[see e.g. ][for an exhaustive review]{kum15}. In the past two decades, the knowledge and understanding of the GRB phenomenology and underlying physics has significantly improved, mainly based on observations by the {\it Neil Gehrels Swift Observatory} ({\it Swift} hereafter; \citealt{geh04}) and {\it Fermi}~\citep[][]{band09} satellites, and on the improved capabilities of ground-based facilities in performing fast follow-up observations. Moreover, a new observational window has recently opened for the study of emission processes in GRBs through the first detections at $\sim$\,TeV energies by different ground-based facilities.

The presence of emission in the GeV-TeV domain in GRBs has been discussed and theorised in several studies (e.g. \citealt{meszaros_rees_94,meszaros_rees_97,zhang_meszaros_00,sariesin}) well before ground-based facilities started to perform follow-up observations at these energies. In order to probe the TeV domain in GRBs, Imaging Atmospheric Cherenkov Telescopes (IACTs), such as the MAGIC telescope, have made a huge effort over the years to become ever more suitable for GRB observations. In particular, efforts were focused on: i) developing a fast repositioning system (able to move at about 7 deg per second) to promptly react to GRB alerts and start observations with delays of a few tens of seconds, and ii) lowering the energy threshold down to 50\,GeV (in optimal conditions) to reduce the impact of the optical/IR photons of the diffuse extragalactic background light~\citep[EBL; ][]{nik61,gou66} on the detection probability of cosmological GRBs. 

In parallel, observations performed with the LAT instrument on board the \textit{Fermi} satellite definitely proved that GeV emission (typically below 10\,GeV) is produced and is a relatively common feature in the brightest events detected by the GBM~\citep[see e.g., ][]{aje19}, the other instrument of the \textit{Fermi} satellite. In the majority of LAT-detected GRBs, GeV emission starts with a short delay (seconds) and lasts much longer (hundreds or thousands of seconds) with respect to the sub-MeV prompt emission. In the time window of overlap between prompt (sub-) MeV emission and LAT emission, a variety of spectral behaviours were found. Sometimes, the LAT spectrum is consistent with being the extrapolation of the prompt keV-MeV spectra \citep{090217A}, while in other cases, a second harder spectral component is clearly detected and is usually well described as an additional power law (PL) that overlaps the Band function \citep{band93} , which describes the prompt emission spectrum and dominates the GBM band~\citep[see e.g., ][]{ack10}. It is still debated whether this second component in the GeV band belongs to the prompt phase or is instead the beginning of the afterglow phase. 
LAT emission at later times is usually well described in terms of afterglow radiation and, in particular, synchrotron emission from electrons accelerated in forward shock \citep{kumar10,ghisellini10,ghirlanda10,beniamini15}.
In the notorious case of GRB~130427A, however, the late arrival (about four minutes after the GRB onset, i.e.\ well after the end of the prompt emission) of a nearly 100\,GeV photon opened the concrete possibility that a component other than synchrotron emission is present (e.g. inverse-Compton or hadronic emission) in the afterglow phase of GRBs and might extend to the very high energy (VHE; $>$\,100\,GeV) domain.
Theoretical expectations in the TeV domain and the complex observational scenario described in the GeV band motivate GRB observations above 100\,GeV, which have constituted one of the most ambitious targets for the current generation of IACTs. 

The MAGIC Collaboration identified the detection of VHE emission from GRBs as one of its primary targets. This key observing program resulted in the very much awaited detection of the first GRB at VHE gamma rays, namely GRB 190114C \citep{MAGIC-190114C-a}, which led to the discovery of an additional spectral component that was detected up to 1\,TeV and was successfully interpreted as synchrotron self-Compton (SSC) afterglow emission \citep{MAGIC-190114C-b}.

Additional detections have enlarged the sample of VHE GRBs (see \citealt{miceli22} for a review). The sample currently consists of five events (all long GRBs, and all detected during the afterglow phase). In addition to GRB~190114C, MAGIC detected VHE gamma-rays (70-200\,GeV from 56\,s to $\sim$2400\,s) from GRB~201216C \citep{grb201216c}, and H.E.S.S.  detected VHE emission from GRB~180720B (100-440\,GeV at $\sim$11\,h; \citealt{grb180720B}) and GRB~190829A (up to 3.3\,TeV; from 4\, to 57\,h, \citealt{grb190829A}). Moreover, LHAASO detected afterglow radiation up to $\sim$\,13\,TeV from GRB~221009A that lasted for $3000$\,s \citep{221009A_lhaaso,221009A_lhaaso_13tev}.
In all these cases, all the VHE data can be successfully described within the SSC scenario (see e.g. \citealt[][]{MAGIC-190114C-b,grb201216c,wang19,salafia22,ren24} and carries a luminosity similar to that of the synchrotron component.

The two GRBs detected by MAGIC are the result of the intense GRB follow-up programme that led to the observation of a considerable number of GRBs over the past years. Several results about GRB follow-up were reported by MAGIC~\citep[][]{alb07,alek10,alek14} and other IACTs~\citep[see e.g., ][]{ver11}. In this paper, we report the results for GRB observations achieved by MAGIC after the implementation of an improved GRB automatic observation procedure in 2013 and before the changes that were introduced after the detection of GRB~190114C. 
The procedure introduced in 2013 allowed us to start the observation within few tens of seconds from the GRB onset. 

For the vast majority of observed GRBs, no gamma-ray signal was detected, and hence, only upper limits (ULs) on the emission flux level could be derived. In this paper, we present these ULs and discuss them within the framework of the few VHE GRBs detected so far, with the aid of the available simultaneous X-ray data. This discussion is conducted to understand how common VHE is in afterglow emission and to constrain its brightness compared to the synchrotron component. This study can be beneficial for the organization of GRB follow-up strategies for current and future IACTs.

The paper is structured as follows: Sect. \ref{sec:magic_grb} introduces the MAGIC instrument and describes the GRB follow-up strategy. Sect. \ref{sec:grbsample} describes the properties of the MAGIC GRB sample and the main selection criteria. Sect. \ref{sec:analysis} describes the data analysis we performed, and Sect. \ref{sec:interesting_grbs} focuses on a multi-wavelength view of a subsample of GRBs observed by MAGIC, in partic-
ular comparing the results with X-ray data. Finally, we discuss and summarise the main results in Sect. \ref{sec:conclusion}.

\section{The MAGIC telescopes and the GRB follow-up}
\label{sec:magic_grb}
The MAGIC system consists of two 17 m dish IACTs located at the Roque de los Muchachos observatory ($28.8^{\circ}$~N, $17.9^{\circ}$~W, 2200 m a.s.l.) on the Canary Island of La Palma, Spain. The MAGIC system is currently carrying out stereoscopic observations with a sensitivity of $ < 0.7 \%$ of the Crab Nebula flux for energies above $\sim 220$ GeV in 50 h of observation, and a trigger energy threshold of 50 GeV \citep[][]{alek16}. Since the beginning of its operation, MAGIC was able to react to GRB alerts through a dedicated automatic alert system (AAS) that receives external triggers provided by the Gamma-ray Coordinates Network (GCN\footnote{\protect\url{http://gcn.gsfc.nasa.gov/}}) through a TCP/IP socket connection. GRB observations are assigned the highest priority within MAGIC operations as soon as the alert is received and validated by the AAS according to predefined criteria. The event coordinates are communicated to the telescope central control, which stops the ongoing observation and starts the re-pointing to the GRB position. Their lightweight structure based on carbon fibre tubes enables the MAGIC telescopes to slew to a target using dedicated fast-slewing movements within few tens of seconds after the alert is received. This implies a remarkable re-pointing speed of about 7 deg per second in both zenith and azimuth\footnote{The normal pointing speed for MAGIC is about 4 deg per second}. To prevent possible failures and problems during the crucial phases of the follow-up of GRBs, an updated automatic procedure was implemented at the beginning of 2013. In particular, within the updated automatic procedure, the data-acquisition system is not stopped during the slewing of the telescopes, and a rate limiter avoids its saturation. This drastically reduces possible failures when the observation of GRBs is started (see details in \citealt{MAGIC-190114C-a}). Depending on the observational constraints, the follow-up of GRBs is usually carried out for a maximum observation time of 4 hours after the beginning of the visibility window. This observation time can be shortened or increased by the burst advocate on shift, who is informed by the local operators and decides the observational strategy based on incoming information on the GRB.

Most of the GRB follow-ups performed by MAGIC are triggered by alerts sent by the BAT instrument on board the \textit{Swift} satellite. While GBM alerts are more frequent, most of them are discarded either due to their large localisation uncertainty (several degrees) or their low significance. This filter is implemented within the AAS in order to avoid the follow-up of events that are less likely to be classified as a real GRB (e.g. terrestrial gamma-ray flashes or cosmic rays) to save valuable observation time. BAT alerts instead provide a very good localisation at the arcminute level, and the probability that the trigger is of non-GRB origin (e.g. a background fluctuation or a flare from a Galactic source) is very low. An even lower number of follow-ups resulted from alerts by the INTEGRAL satellite \citep{integral} because this instrument detects intrinsically few GRBs. The localisation is good enough to avoid adding additional filters in this case as  well. Finally, a non-negligible number of follow-ups was triggered by GRBs detected by LAT, either as a follow-up of GBM detected bursts that were afterwards also detected by LAT, or as a late-time observation. In the latter case, the follow-up is performed even more than one day after the GRB onset. This observation strategy for LAT-detected alerts has a twofold explanation: on the one hand, it allows us to test for the presence of long-lasting emission even hours after the GRB onset, and on the other hand, if the GRB is detected by GBM but not by BAT and is initially discarded, it allows us to search for a possible gamma-ray signal from a GRB exhibiting GeV and possibly VHE emission. 

To quantify the duty cycle of MAGIC (or of any Cherenkov telescope located at the Roque de los Muchachos) for the observation of GRBs and to cross-check for systematic errors that can be nested in the follow-up procedure, we present here a calculation of the expected percentage of successfully followed alerts. We start our calculation from the constraints that are included in the automatic system that limit the number of alerts that can be followed. The criteria are listed below.
\begin{itemize}
\item The Sun must be below the astronomical horizon (103\,deg from zenith) to reduce the available time by 39\% .
\item Alerts are followed if within 4 hours from the trigger time the position in the sky is 60\,deg from zenith at most. This further reduces all observable alerts by 37.5\% (25\% of the alerts would have a zenith angle below 60 deg at the trigger time, and an additional 12.5\% would satisfy the zenith requirement within 4 hours). 
\item When the Moon is above the horizon, the angle between the GRB and the Moon must be larger than 30 deg in order to avoid pointing too close to the Moon, which would lead to a high-energy threshold and increased systematics. This has a small impact and excludes only 3.4\% of the observable alerts.
\end{itemize}
Moreover, we need to take into account the days in a year in which MAGIC does not perform GRB follow-up because operations are stopped due to full-Moon conditions. These sum up to 21\% of the total available night-time throughout the year.
Finally, we need to account for the fraction of time in which operations are stopped due to weather above the operational safety limits or technical problems, which is 40\% on average. 
When all the factors listed above are considered, the percentage of alerts that the MAGIC telescopes are expected to follow in a year is 6.7\%

We then applied this percentage to the number of GRB alerts released between 2013 and 2019 and estimated the number of GRBs that MAGIC should have followed. Most alerts are provided by {\it Swift} and {\it Fermi}. To the alerts coming from GBM, an additional requirement on the error on the GRB localization is applied. Only alerts with a statistical error smaller than 1 deg are processed by the automatic system. This additional constraint is necessary because of the large localization uncertainty of GBM in comparison to other instruments (several degrees versus arcminutes as provided e.g.\ by \textit{Swift}-BAT). Given the relatively small field of view of MAGIC (3.5 deg in diameter), we need to maximise the probability that the actual location of the GRB falls within
the effective field of view of the MAGIC camera. The number of GRB alerts that MAGIC is expected to follow up per year in the period 2013-2019 is about 6.7 from {\it Swift} and about 2.9 from {\it Fermi}, which means a total of 9.6 GRBs per year. In the 7 years considered in this manuscript, from 2013 to 2019, MAGIC performed 66 GRB observations, which agrees well with the expected number of observable alerts. We received a total of 46 alerts from {\it Swift} (6.6 per year) and 17 from {\it Fermi} (2.4 per year). The remaining 3 alerts were sent by {\it INTEGRAL}. 

Finally, after several years in which the updated automatic procedure was used and refined, we can assess its reliability. We took into account the number of times it was used after its introduction and the number of times it failed in the case of real alerts. From May 2013 to December 2019, the automatic procedure took over 48 times, and it failed only twice. Therefore, the efficiency of the updated automatic procedure is about 96\%. It has to be noted that the issue that occurred in the two failure cases is not specific to the automatic procedure, but occurs rarely during normal data taking. \\

\section{The sample}
\label{sec:grbsample}

\begin{figure}
\centering
\includegraphics[width=\columnwidth]{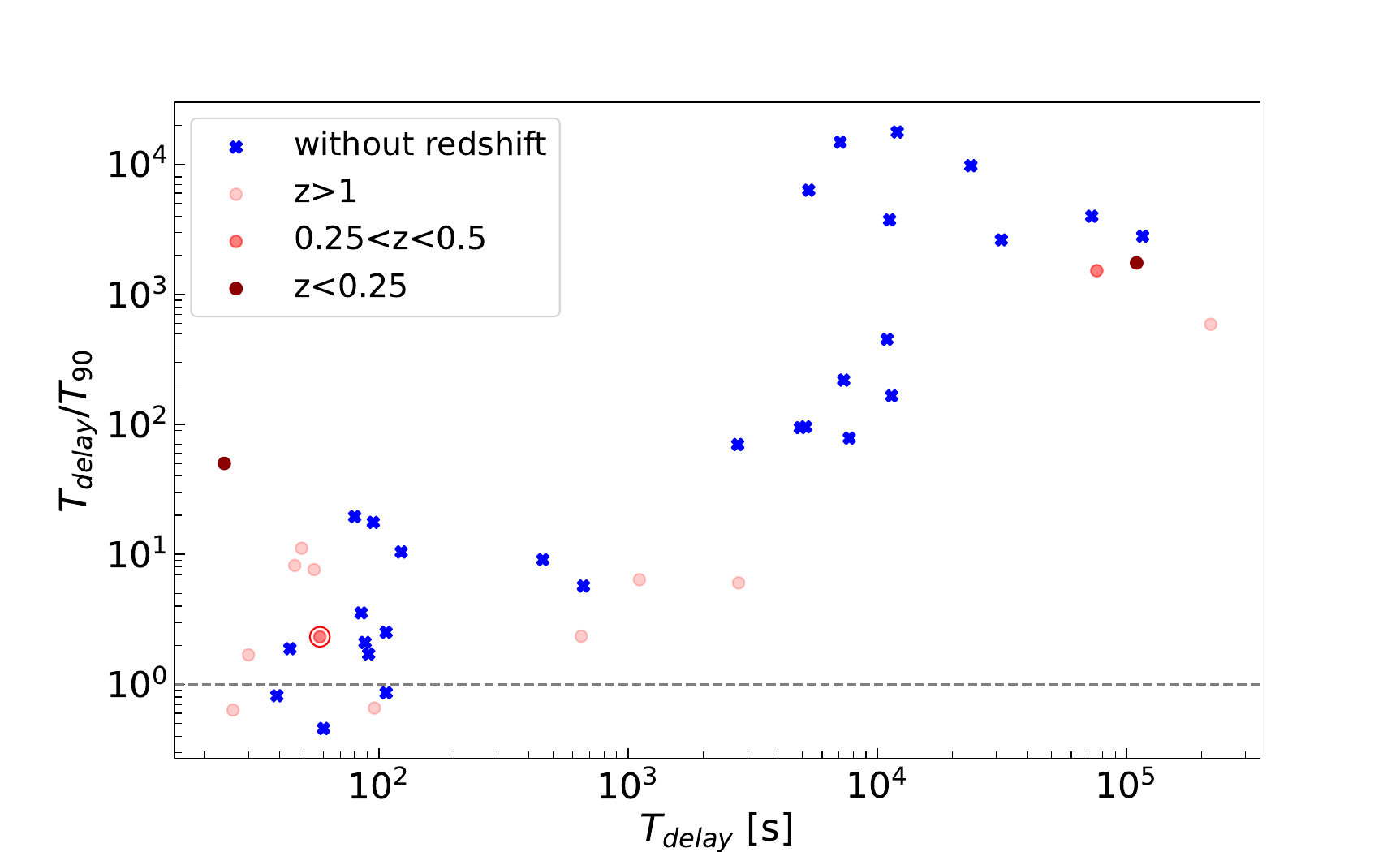}
\caption{Ratio of the observation delay $T_{delay}$ and $T_{90}$ vs. $T_{delay}$ for the GRBs listed in Table \ref{tab:grb}. GRBs with and without redshift are denoted by red and blue markers, respectively. The dashed horizontal line denotes when the ratio is equal to one, meaning that for GRBs with a ratio lower than one ,observations started on a timescale comparable to the duration of the prompt emission. For the sample under consideration, the GRBs fulfilling these conditions are GRB131030A, GRB141026A, GRB150428B, GRB170728B, and GRB180720C. The only GRB from this sample that was significantly detected at TeV energies, namely GRB~190114C, is marked with an empty circle.}
\label{fig:delay_t90_ratio}
\end{figure}

\begin{figure}
\centering
\includegraphics[width=\columnwidth]{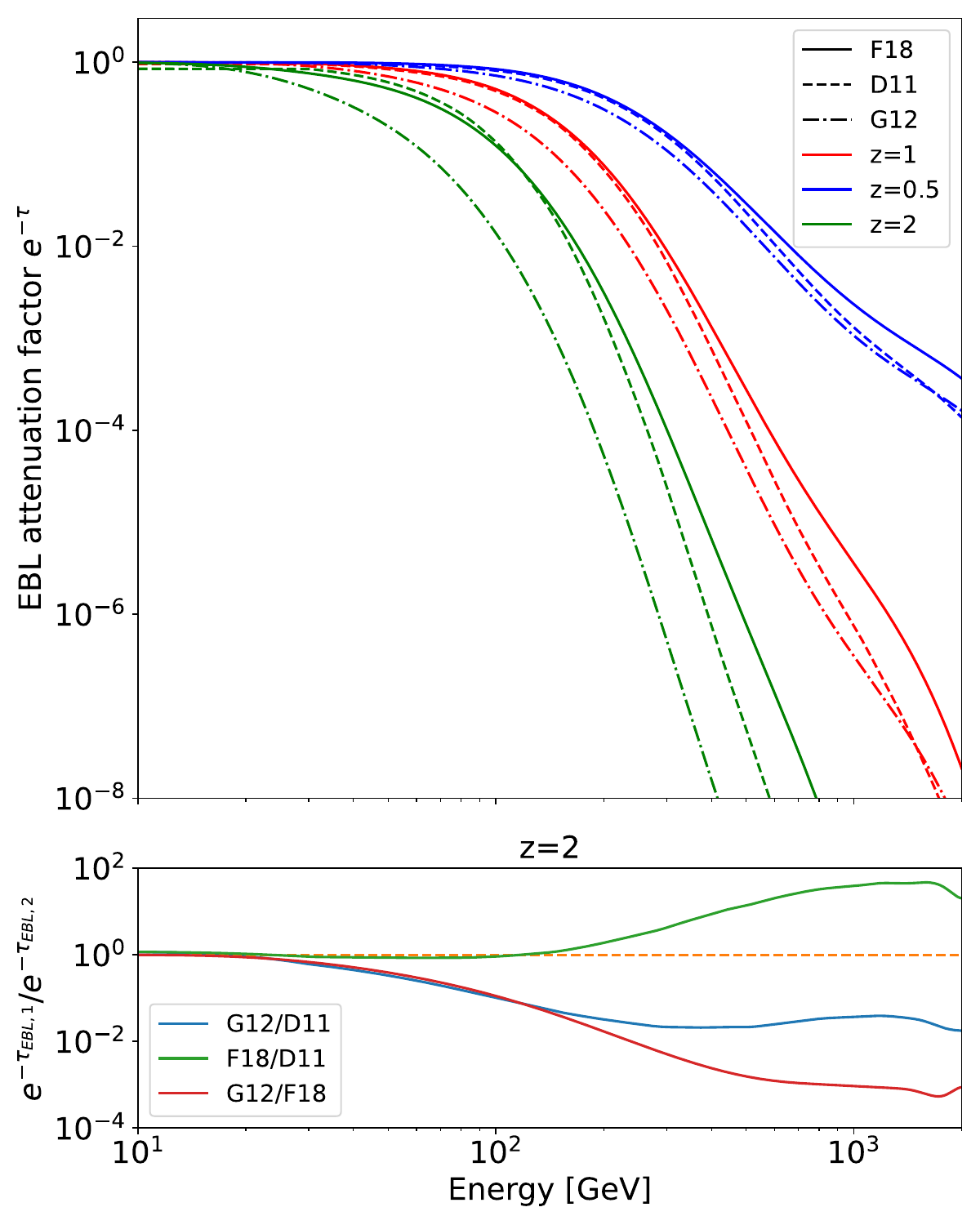}
\caption{ Top panel: EBL attenuation factor in the 0.1 - 10 TeV energy range for three different redshift values (z = 0.5, 1, and 2) and for the three different EBL models we used in this study: \cite{Dominguez} (D11), \cite{ebl_franceschini_2018} (F18), and \cite{ebl_gilmore_2012} (G12). Bottom panel: Ratio of the EBL attenuation factors for the three EBL models adopted in this study (D11, G12, and F18) at redshift z = 2.} 
\label{fig:ebl_tau}
\end{figure}

We focus on GRBs that were observed by MAGIC in the time period from 2013 to 2019. The choice of the starting year was motivated by the fact that at the beginning of 2013 the automatic procedure for GRB follow-up was updated. In particular, the upgrade aimed to reduce possible hardware failures during and after the fast slewing of the telescopes and to change the GRB observation mode from on-source to wobble pointing \citep{fomin94}. The latter change also led to a simplification of GRB data analysis, which now makes use of standard tools developed within the MAGIC collaboration (see Sect. \ref{sec:analysis}). 
The choice to stop in 2019 was motivated by the revision of the strategy for GRB follow-up after the detection of GRB~190114C. Until 2019, late-time follow-up observations were performed only in case of a LAT detection. However, the connection seen between X-ray and VHE band in GRB\,190114C indicates that X-ray emission can also be a reliable proxy for the identification of potentially interesting events in the GeV-TeV domain. The following GRB detections in the VHE domain  confirmed the validity of this approach. As a result, from 2020 on, several late-time follow-ups were performed even in cases without LAT detection. 
The chosen period of investigation of this study assures that the sample is collected under homogeneous selection criteria and observing strategy.

MAGIC followed-up a total of 66 GRBs in the selected period. We discarded 24 follow-up events from the analysis because the data were taken in conditions of degraded or unstable atmospheric transmission (21 out of 24) or because technical issues severely affected the quality of the data acquisition (3 out of 24) .

The resulting sample included 42 GRBs (38 long and 4 short, according on the classification based on $T_{90}$ being longer or shorter than 2 seconds). Out of these 42 events, 15 have measured redshift (14 long and 1 short). The sample is listed in Table~\ref{tab:grb}, which reports the redshift (when available), the information on which instrument provided the localisation adopted by MAGIC, the duration ($T_{90}$) of the prompt emission, the trigger time $T_0$, the time $T_{\rm start}$ when MAGIC observations started, the delay time of MAGIC observations ($T_{\rm delay} = T_{\rm start} - T_0$), and the range of zenith angles during the observations.

Fig.~\ref{fig:delay_t90_ratio} shows the ratio of the time delay $T_{\rm delay}$ of MAGIC observations and the duration of the prompt emission (identified with the $T_{90}$) versus $T_{\rm delay}$. 
An interesting statistics is the number of times when the observation started while the prompt emission was still ongoing. These cases are roughly identified by the condition $T_{\rm delay}/T_{90}\lesssim1$. In 5 (out of 42) cases, the observations started within the time window defined by $T_{90}$: GRB~131030A, GRB~141026A, GRB~150428B, GRB~170728B, and GRB~180720C. These are the best candidates for detecting or constraining a possible VHE emission associated with the prompt phase. However, we note that for three of them, the redshift is unknown and no meaningful constraints on the intrinsic luminosity can therefore be inferred (see Sect.~\ref{sec:analysis}). The other two GRBs with $T_{\rm delay}/T_{90}<1$ have a high redshift ($z=1.295$ and $z=3.35$), implying a strong absorption of the VHE emission by EBL. Moreover, in GRB~131030A ($z=1.295$). the first observation run was discarded at the offline analysis level because it was heavily affected by clouds above the MAGIC site, and only observations after $\sim10^3$\,s could be exploited. 

In general the best candidates for the detection of VHE emission are those GRBs detected at early times, when the afterglow emission is still bright. However, Fig.~\ref{fig:delay_t90_ratio} shows that most of them have an unknown redshift or redshift greater than one. The only GRBs with $T_{\rm delay}<10^4$\,s and $z<1$ are GRB~190114C (which was indeed detected) and GRB~160821B, which is a short GRB (then with an intrinsically less luminous afterglow). A possible hint of a detection of GRB~160821B was discussed by \cite{160821b}. Other GRBs at redshift $z<0.5$, which might be very interesting for a VHE detection, were followed only after $\sim1$\,d (see Fig.~\ref{fig:delay_t90_ratio}).

\section{Data analysis and flux calculation}
\label{sec:analysis}

\subsection{MAGIC data analysis}
The analysis of GRB~160821B, GRB~190114C, and GRB~190829A are not presented in this article because they have already been extensively presented or will be presented in dedicated papers. GRB~160821B shows a hint of a detection, as discussed by \cite{160821b}. GRB~190114C was clearly detected, and the results of the analysis and interpretation are available in \cite{MAGIC-190114C-a,MAGIC-190114C-b}. A study on the MAGIC observation of GRB~190829A is currently ongoing and a separate publication is foreseen in which a marginal detection is reported.

The data analysis of the remaining 39 GRBs was performed using the standard software package MARS \citep[][]{zan13}. After the calibration of the single events recorded by each telescope \citep[][]{alb08}, the data were processed with a cleaning process \citep[][]{alek16} with 6-3.5 photo-electrons (phes) threshold levels for the core and boundary pixels when data were taken without the presence of the Moon. Higher cleaning levels were adopted for observations performed in the presence of the Moon, as prescribed by \citep{moon_performance_magic}. For the gamma/hadron separation and the reconstruction of the direction of the events, a multivariate method based on a random forest (RF) algorithm was applied~\citep[][]{alb08b}. This algorithm employs some Cherenkov image parameters ~\citep{hil85} calculated for Monte Carlo gamma-ray simulations and real data, containing mostly hadronic events, to compute a gamma/hadron discriminator called \emph{hadronness}, which can take values between zero and one (zero for a gamma event, and one for a hadron event). The energy of the events was estimated by averaging individual energy estimators for the two telescopes based on look-up tables. The possible detection of a gamma-ray signal is investigated through the so-called $\theta^2$ plot, that is, the comparison of the distributions of the squared angular distance between the reconstructed position of the source and its nominal position in the signal and background regions for energies above the threshold. The significance of the signal was evaluated using single cuts in \emph{hadronness} and $\theta^2$ and according to equation~17 of~\cite{lim83}. The performed analysis did not reveal significant (above $5\sigma$) VHE emission for any of the observed GRBs during the MAGIC follow-up. 

\subsection{Energy flux upper limits calculation}
The ULs were evaluated assuming a spectral shape for the intrinsic or the observed differential photon spectrum using the method of \cite{rol05}, with a confidence level (CL) of 95$\%$ and a total systematic uncertainty of 30$\%$. Depending on the available information on the redshift and zenith angle observations, we followed a different approach for the calculation of the ULs to avoid large systematics uncertainties and a strong impact from EBL absorption. The EBL attenuation can reduce the flux in the gamma-ray domain by several orders of magnitude, also providing different results depending on the adopted EBL model (see Fig.~\ref{fig:ebl_tau})

\subsubsection{GRBs with unknown $z$, $z>2$, or Zd $ > 40$ deg} 

For the GRBs in our sample with an unknown redshift or with a redshift $z \geq 2$ or observed at zenith angle Zd $ > 40$ deg, we estimated night-wise ULs on the observed flux. This applied to 33 of the 39 GRBs considered for this analysis.
We did not consider energies above 1\,TeV. For a high redshift, the effect of EBL absorption above 1\,TeV is extreme (see Fig.~\ref{fig:ebl_tau}) and leads to huge uncertainties of the results that would be derived above this energy.

The observed photon spectrum up to 1\,TeV was assumed to be described by a power-law function $dN/dE = E^{-\alpha}$. We considered two possible values for the photon index $\alpha$: 3.5 and 5.5. The first represents an optimistic scenario that is compatible with an event similar to GRB\,190829A, and the second is compatible with the results obtained for GRB\,190114C and GRB\,201216C. The resulting ULs on the observed flux in several energy bins and time intervals are listed in Table \ref{tab:observed_uls} in the appendix for this subsample of 33 GRBs . Only energy bins that ensured a systematic uncertainty below 30$\%$ are reported. In addition, the time window in which the ULs were computed was selected by excluding intervals of bad atmospheric conditions or problems during the data taking. As a result, the starting time for the UL calculation (T$_{start}$ in Table \ref{tab:observed_uls}) may differ with the time at which the telescopes started the follow-up observations ($T_{delay}$ reported in Table \ref{tab:grb}). For three GRBs of the sample, namely GRB160910A, GRB190106B and GRB191004A, it was not possible to derive any ULs because none of the energy bins fulfilled the criteria for the systematics and the EBL absorption.

\begin{figure*}
\centering
\includegraphics[width=\columnwidth]{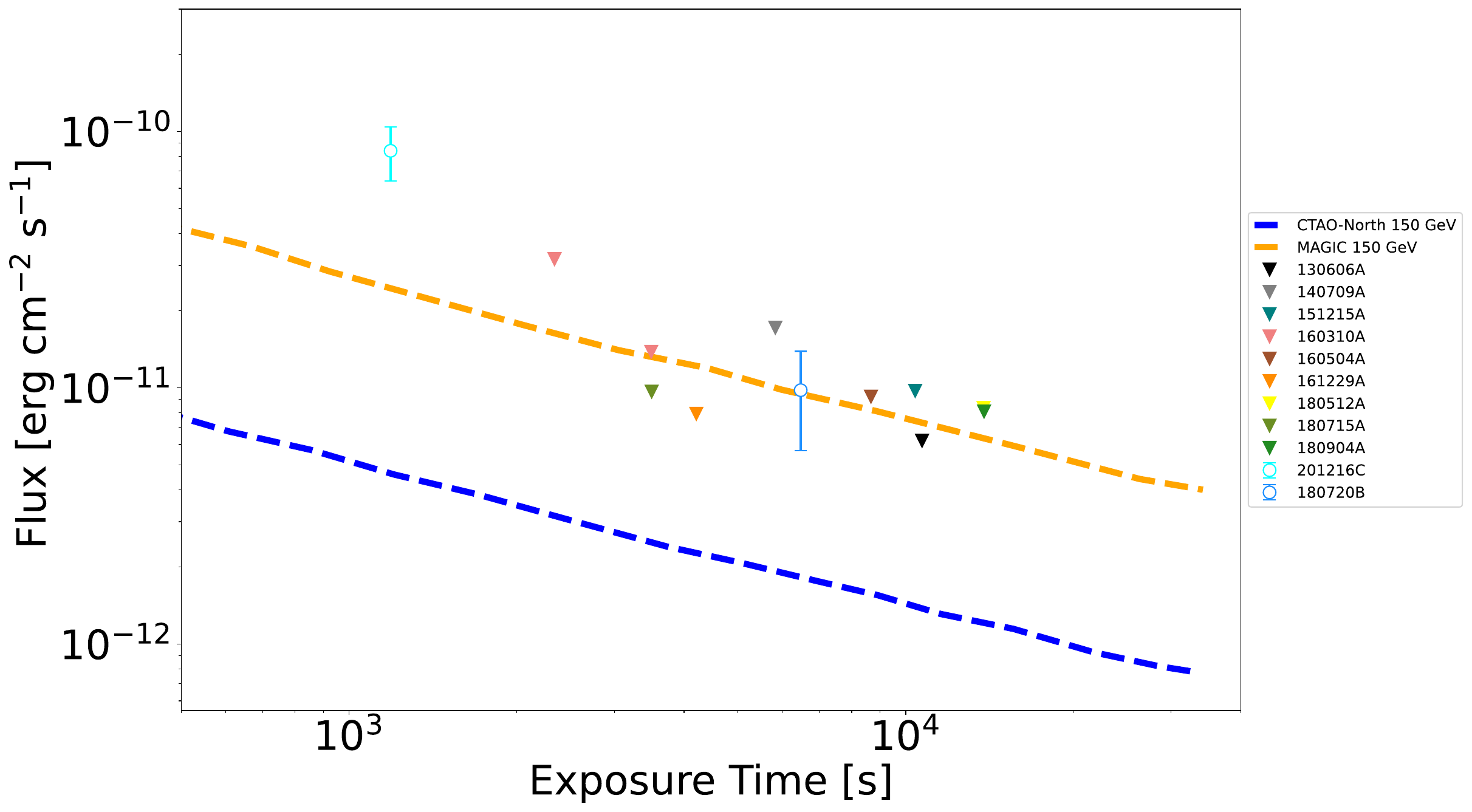} 
\includegraphics[width=\columnwidth]{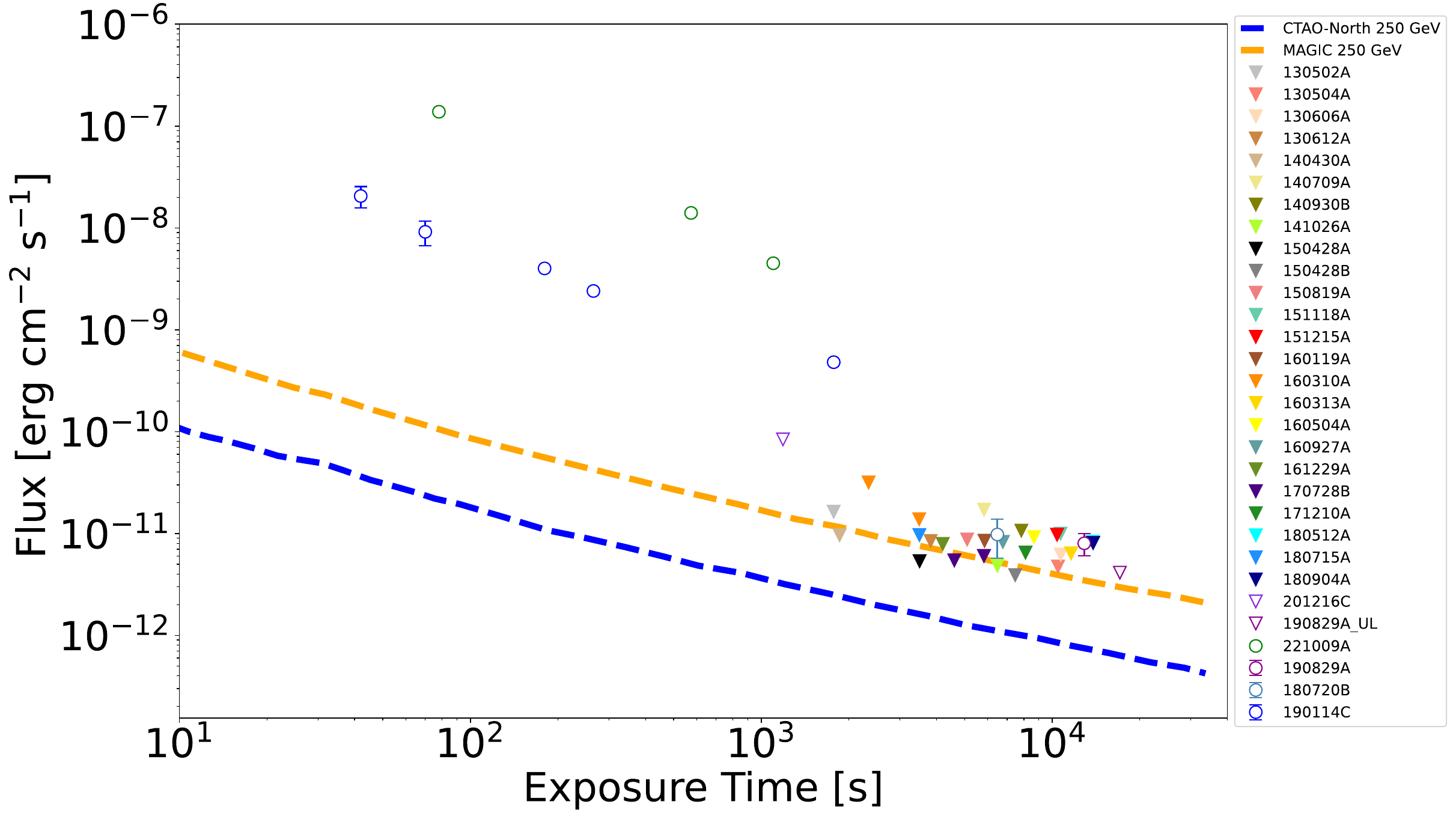}
\caption{Observed flux points or ULs for TeV-detected GRBs (empty markers) and most stringent ULs for catalog of GRBs non-detected by MAGIC (filled markers) vs. exposure time. The $2\sigma$ sensitivity level curves of the MAGIC and CTAO-North array are also displayed. Two reference energy values are shown: 150 GeV (left plot) and 250 GeV (right plot).}
\label{fig:non_interesting_grbs}
\end{figure*}

The energy flux ULs \footnote{From now on, we refer to energy flux ULs simply as flux ULs} estimated at 150\,GeV and 250\,GeV are plotted as a function of the exposure time in Fig.~\ref{fig:non_interesting_grbs}. The two considered reference energy values (150 GeV and 250 GeV) allowed us to be less sensible to EBL absorption and to include most of the GRBs of our sample. For comparison, we also report the fluxes or ULs for the TeV-detected GRBs in their corresponding exposure times. The flux values and corresponding time exposures were extracted from the papers of the TeV-detected bursts by MAGIC, HESS, and LHAASO. In a few cases (GRB190829A and GRB201216C), we estimated only an UL, rather than a flux point, for the selected reference energy value. We also plot the sensitivities at a $2 \sigma$ level of the MAGIC and the Cherenkov Telescope Array Observatory (CTAO) northern array. The sensitivities at the $2 \sigma$ level for the CTAO-North telescopes were derived from the official CTAO web page\footnote{\url{https://www.cta-observatory.org/science/ctao-performance/\#1472563157332-1ef9e83d-426c}}, while those for MAGIC were derived from \cite{fioretti}. The sensitivity curves at a $5\sigma$ level were rescaled to those at a $2\sigma$ level assuming that the sensitivity, $S,$ scales with significance $\sigma$ as $S(2\sigma)/S(5\sigma)$ $\propto$ $2/5$. This is a simplified approach, but it can be provide a valid estimate for a background-dominated regime \citep{ambrogi}.

\subsubsection{GRBs with $z < 2$ and Zd $< 40$ deg} 
\label{subsubsec:interesting_grbs}

\begin{table*}[!ht]
\begin{center}
\caption{\centering ULs at the 95\% confidence level on the intrinsic flux (integrated from $E_{\rm min}$ to $E_{\max}$) for the subsample of GRBs with $z<2$ and Zd $ < 40$ deg.}
\begin{tabular}{cccc|cccc}
\toprule
GRB    & T$_\textup{obs}$ & $E_{\rm min}$& $E_{\rm max}$ & F18,1.6 & F18,2.2 & G12,1.6 & G12,2.2 \\
\scriptsize{}      & \scriptsize{} & \scriptsize{} & \scriptsize{} & \tiny{10$^{-10}$}& \tiny{10$^{-10}$}& \tiny{10$^{-10}$} & \tiny{10$^{-10}$}\\
     \text{name}      & \text{[s]} & \text{[GeV]} & \text{[GeV]} & \tiny{[erg\,cm$^{-2}$\,s$^{-1}$]}& \tiny{[erg\,cm$^{-2}$\,s$^{-1}$]}& \tiny{[erg\,cm$^{-2}$\,s$^{-1}$]} & \tiny{[erg\,cm$^{-2}$\,s$^{-1}$]}\\
\midrule
130701A &  403    & \num{100}  & \num{696} &  4.45   &   2.70       & 11.4 &	6.57 \\
130701A &  1935    & \num{100}  & \num{696} &  1.76   &   1.06       & 4.45 &	2.55 \\
131030A &  5795	   & \num{120} & \num{654} &  1.64	    &    1.00      & 5.72   & 3.30 \\
141220A &  314    & \num{75}  & \num{647}&  4.10	    &     2.39     & 10.7   &	5.85	\\
141220A &  2386    & \num{75}  & \num{647}&  1.92	    &     1.11     & 4.97   &	2.70	\\
160623A &  9324	   &  \num{165}   &  \num{1097}&  	0.56    &    0.40      & 0.71   & 0.50	\\
160623A &  8388	   &  \num{140}        &  \num{1097}&  	0.14    &    0.10      & 0.22   & 0.12	\\
160625B &  13968    &  \num{200}  &  \num{625}  &  7.58	    &    5.11      & 56.3   & 35.9	\\
160625B &  8100    &  \num{110} &  \num{625}  &  1.54	    & 0.95          &  5.94  &	3.45 \\
171020A &  942    &  \num{110} &  \num{523}  &  5.36	    & 3.20          &  36.7  &	19.9 \\
171020A &  11906    &  \num{110} &  \num{523}  &  5.42	    & 3.32          &  41.1  &	23.3 \\
\bottomrule
\end{tabular}
\end{center}
\tablefoot{Flux ULs are  listed in units of $10^{-10}$\,erg\,cm$^{-2}$\,s$^{-1}$. The abbreviations refer to the EBL models (F18 and G12; see text) and to the assumed intrinsic photon indices (1.6 and 2.2). The table also lists the observation time and the lower and upper energy edge for the UL calculations. Multiple entries for the GRBs refer to different time intervals, as described in Sec.~\ref{subsubsec:interesting_grbs}.}
\label{tab:interesting_uls_grb}
\end{table*}

For the GRBs in the sample with a redshift $z < 2$ and a zenith angle Zd $ < 40$ deg, we estimated de-absorbed (i.e. corrected for the EBL absorption) ULs in selected energy and time intervals assuming a power-law function for the intrinsic gamma-ray differential photon spectrum $dN/dE = E^{-\alpha}$. This applied to 6 of the 39 GRBs considered for this analysis. Two values of the photon index $\alpha$ were considered: 1.6 and 2.2. This choice can be justified from a phenomenological point of view because the values correspond to the best-fit values obtained for GRB~180720B and GRB~190114C \citep{HESS19,MAGIC-190114C-a}. Moreover, they are also close to the photon indices expected in the SSC scenario: if the radiation is generated by electrons that are accelerated and result in a power-law distribution $dN/d\gamma$ = $\gamma^{-p}$ with $p$ ranging between 2.2 and 2.4, the photon indices $\alpha=1.6$ and $\alpha=2.2$ roughly correspond to the photon indices of the SSC spectrum assuming that the peak is above or below the GeV-TeV band. To estimate the effects of EBL absorption, we considered three different EBL models: \cite{Dominguez} (D11), \cite{ebl_franceschini_2018} (F18), and \cite{ebl_gilmore_2012} (G12). The ULs computed with the EBL models D11 and F18 differ by no more than 30$\%$ up to $E \sim 200$ GeV (see the bottom panel of Fig.~\ref{fig:ebl_tau}) , and we therefore decided to report only the latter ones. With these assumptions for each GRB, we calculated four values of the ULs in each energy bin. The results are reported in Table~\ref{tab:interesting_uls_grb}. In the following paragraphs, we explain some considerations related to the energy intervals and the time windows reported in Table~\ref{tab:interesting_uls_grb}.

\noindent \textit{Energy intervals}: The low-energy edge of the flux-integration window $E_{min}$ is given by the energy threshold of each GRB, calculated as the peak of the MC reconstructed energy distribution weighted for the observed spectrum, or by the lowest-energy value that ensures a total systematic uncertainty within 30$\%$. In order to assess this condition, the corresponding estimated effective area in the selected energy range was evaluated for different values of $E_{min}$ starting from the GRB energy threshold. 
Given a value of $E_{min}$, the estimated effective area 
was calculated assuming $E_{min}$ or a value shifted by $\pm$ 15$\%$ as the low-energy edge of the flux-integration window. 
The estimated effective area calculated for the shifted energy edges must vary by less than 30$\%$ with respect
to the one computed for the nominal chosen $E_{min}$. If the GRB energy threshold did not fulfill this condition,
an increased energy value that ensures that the variation in the total effective area was 
lower than 30$\%$ was chosen for the UL calculation. The upper-energy edge $E_{max}$ was fixed to 1.5\,TeV in the rest frame. This value was chosen considering that the highest photon energies observed by MAGIC from GRB~190114C are $\sim1\,$TeV (observer frame, corresponding to $\sim1.5$\,TeV in the rest frame). 

\noindent \textit{Time intervals}: For three GRBs of the subsample, namely GRB\,130701A, GRB\,141220A, and GRB\,171020A, the MAGIC observation started with a short delay (shorter than $\sim 100$ s) and covered several hours. Considering the typical rapid evolution of the afterglow flux, especially in the first hours after the trigger time, a single UL value cannot provide enough relevant information to be used in multi-wavelength studies, as we show in the following section. For this reason, we decided to estimate the ULs by splitting the total observational interval into two time bins. For GRB\,160623A and GRB\,160625B, late-time observations were performed for two nights. For this reason, the analyses were performed separately, and two different sets of ULs were computed for each night. For GRB~160625B, during the first night, in particular. from $T_0+ 2769$ s to $T_0+ 8575$ s, although a MAGIC observation was performed, it was not possible to derive ULs that fulfilled the criteria for the total systematic uncertainty explained above due to a combination of medium to large zenith angle (40 deg < Zd < 60 deg) and high night-sky background conditions due to the presence of the Moon. For this reason, we decided to focus on the data collected at zenith below 40 deg that were collected starting from $T_0+ 8575$ s.
For GRB\,131030A, we removed the data that were collected in the first 20 minutes of observations from $T_0+ 27$ s to $T_0+ 1260$ s because they were strongly affected by poor atmosphere conditions.

\section{Comparison with X-ray fluxes}
\label{sec:interesting_grbs}

\begin{figure*}
\centering
\includegraphics[width=0.48\textwidth]{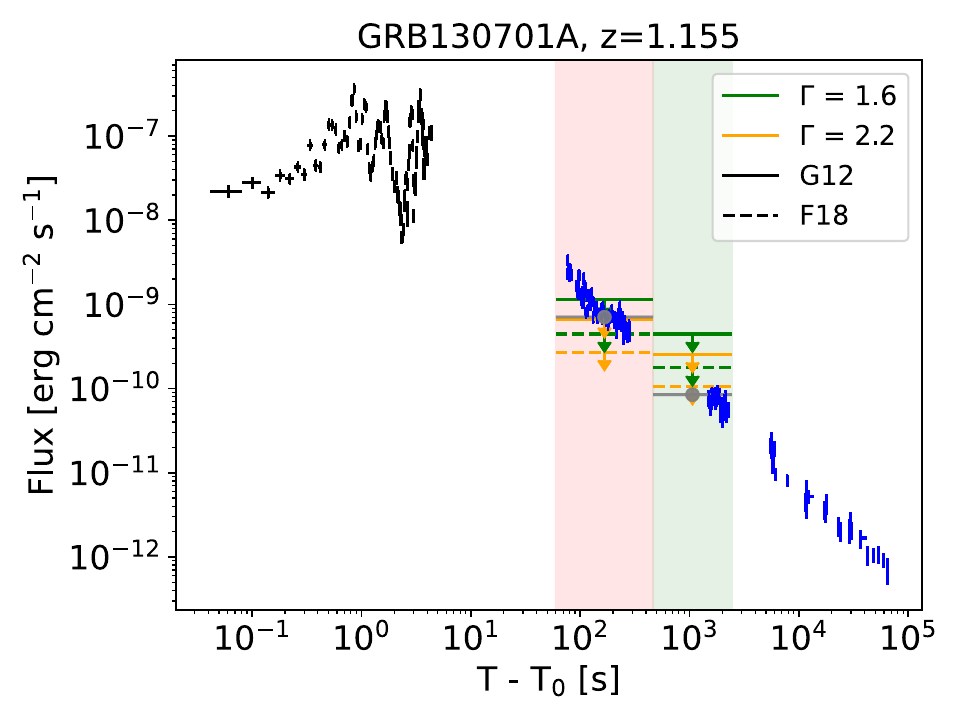} 
\includegraphics[width=0.48\textwidth]{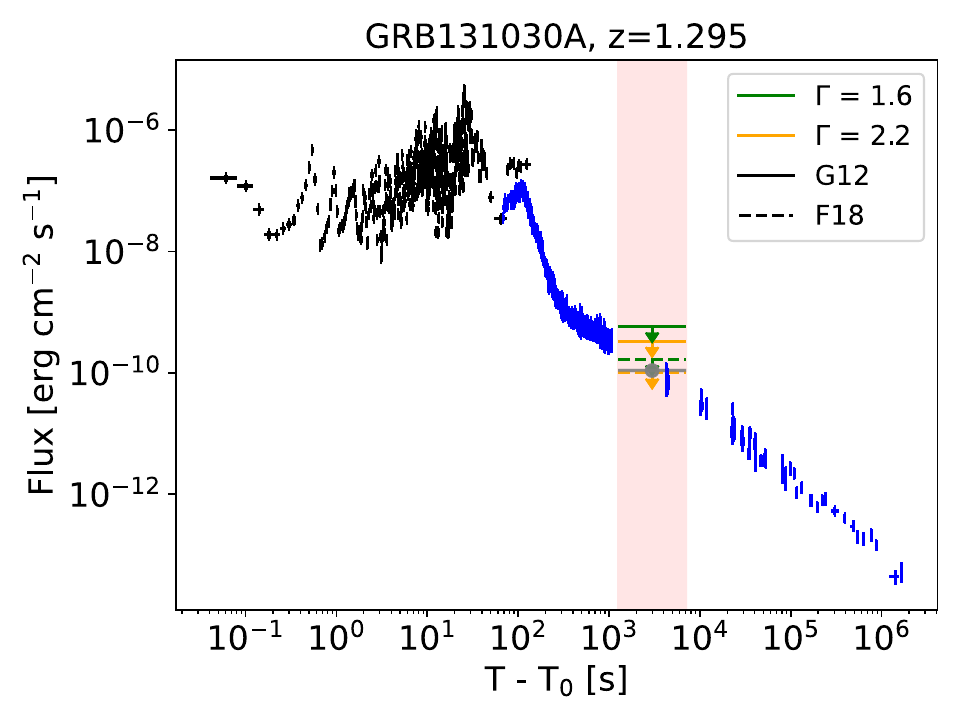}
\includegraphics[width=0.48\textwidth]{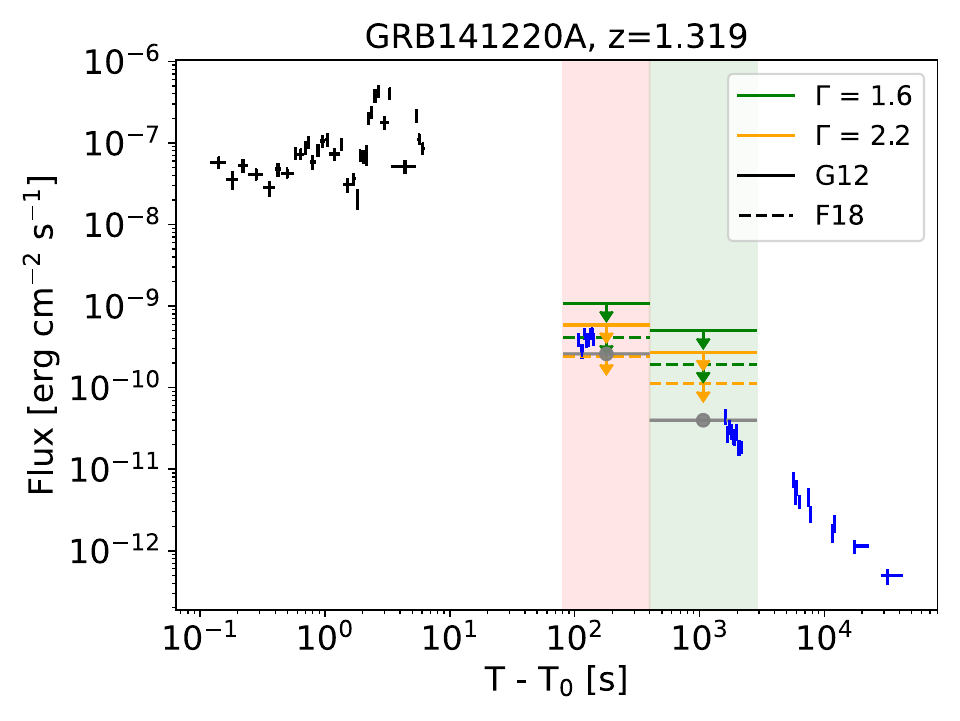}
\includegraphics[width=0.48\textwidth]{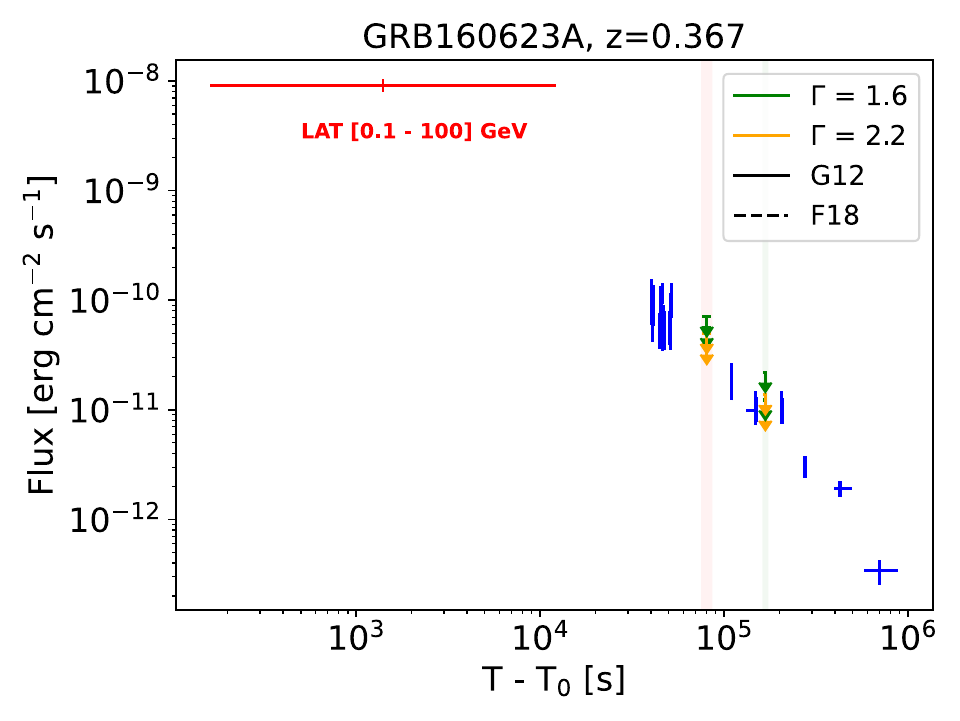}
\includegraphics[width=0.48\textwidth]{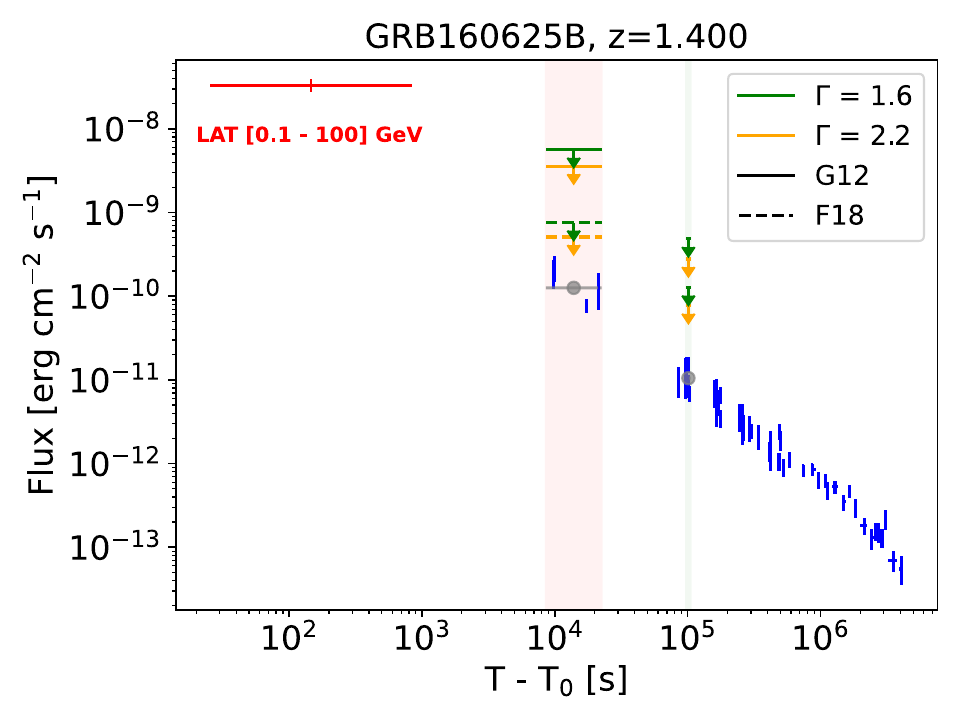}
\includegraphics[width=0.48\textwidth]{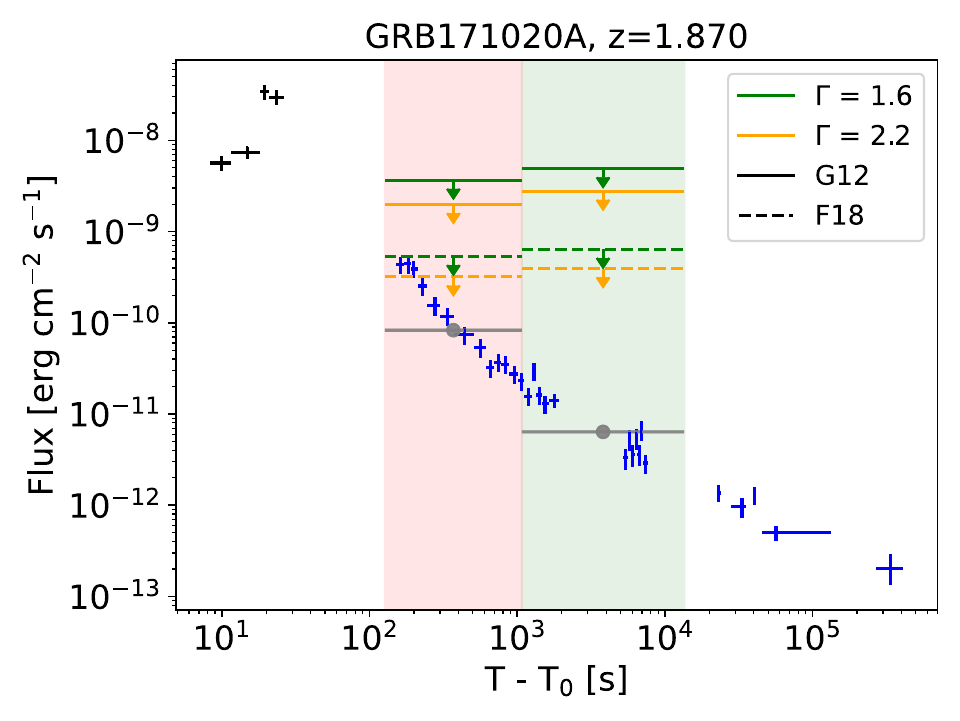}
\caption{Multi-wavelength light curves of the subsample of six GRBs described in Sect. \ref{subsubsec:interesting_grbs} and Table \ref{tab:interesting_uls_grb}. We show the flux light curves with X-ray data (black for BAT and blue for XRT), average X-ray flux in the MAGIC observational time windows (grey points), LAT data (red, if present), and MAGIC ULs assuming two different photon indices and EBL models for the subsample selected for the comparison with lower-energy bands. The time windows in which MAGIC ULs were computed are marked with vertical red and green stripes. }
\label{fig:interesting_grbs}
\end{figure*}

The current population of GRBs detected at VHE share one common behaviour: the simultaneous EBL-corrected luminosities in the soft X-ray band and in the VHE band are comparable. 
In an SSC scenario, this roughly implies a similar amount of power in the synchrotron and in the inverse Compton components, assuming that the peak of the synchrotron component is approximately the soft X-ray band and the peak of the SSC is not far from the VHE band (see e.g. Fig. 2 and 3 in \citealt{MAGIC-190114C-b}).
Following these considerations, the question immediately arises: do all GRBs have a VHE emission component with a luminosity similar to the simultaneous X-ray luminosity) If this is not the case, are the VHE-detected GRBs a peculiar population with particularly bright VHE emission? We partially investigated these open issues making use of the MAGIC ULs. Our aim was to understand how the ULs on the intrinsic VHE luminosity compare to simultaneous X-ray luminosities.
We therefore compared the MAGIC de-absorbed flux ULs and the de-absorbed (i.e. corrected for Galactic and intrinsic absorption) flux in the soft X-ray band for a subsample of GRBs. For this comparison, we selected GRBs with a measured redshift, limiting the sample to redshift $z < 2$, and for which the total systematic uncertainty of the MAGIC ULs in the selected energy range was below 30$\%$. For $z > 2$ and/or for energies above hundreds of GeV, the effect of EBL is so strong that the de-absorbed fluxes are much higher than the X-ray fluxes, and they do not provide meaningful constraints. 

For the selected sample, we plot the de-absorbed flux light-curve integrated in the energy range 0.3-10\,keV (observer frame, from \url{https://www.swift.ac.uk/xrt\_curves/}) from the XRT instrument on board the \textit{Swift} satellite together with MAGIC ULs on the intrinsic flux (i.e. de-absorbed by the EBL attenuation). 
The energy range over which the MAGIC ULs were calculated varies among the selected GRBs and is listed in Table ~\ref{tab:interesting_uls_grb}.
Since the XRT data points are typically derived on much shorter timescales, for a more reliable comparison, we estimated the XRT average flux over the same time interval where the MAGIC flux UL was extracted. To do this, we first fitted the X-ray light curves with a power-law function and then used the best fit to infer the average flux in the MAGIC time window. 
The results are shown in Fig.~\ref{fig:interesting_grbs}. 
The XRT light curves are shown with blue data points, while the grey filled circles mark the XRT fluxes averaged over the same time-window of the MAGIC observations. In some cases, the total temporal window of MAGIC observations has been divided into two different time bins, for a more meaningful comparison. Arrows show the flux UL inferred from MAGIC observations. Four different estimates of the ULs are provided, inferred from two different assumptions on the photon index of the intrinsic spectrum and for two different EBL models (see the legend).
For GRB~160623A and GRB~160625B, LAT observations are available, but they unfortunately do not overlap in time with MAGIC observations. We added the estimated flux in the $0.1 -100$ GeV band as calculated by LAT \cite{aje19} (see the red data points in Fig.~\ref{fig:interesting_grbs}). For each GRB displayed in Fig.~\ref{fig:interesting_grbs}, we also included BAT data for completeness (black data points).

\begin{figure*}
\centering
\includegraphics[width=0.48\textwidth]{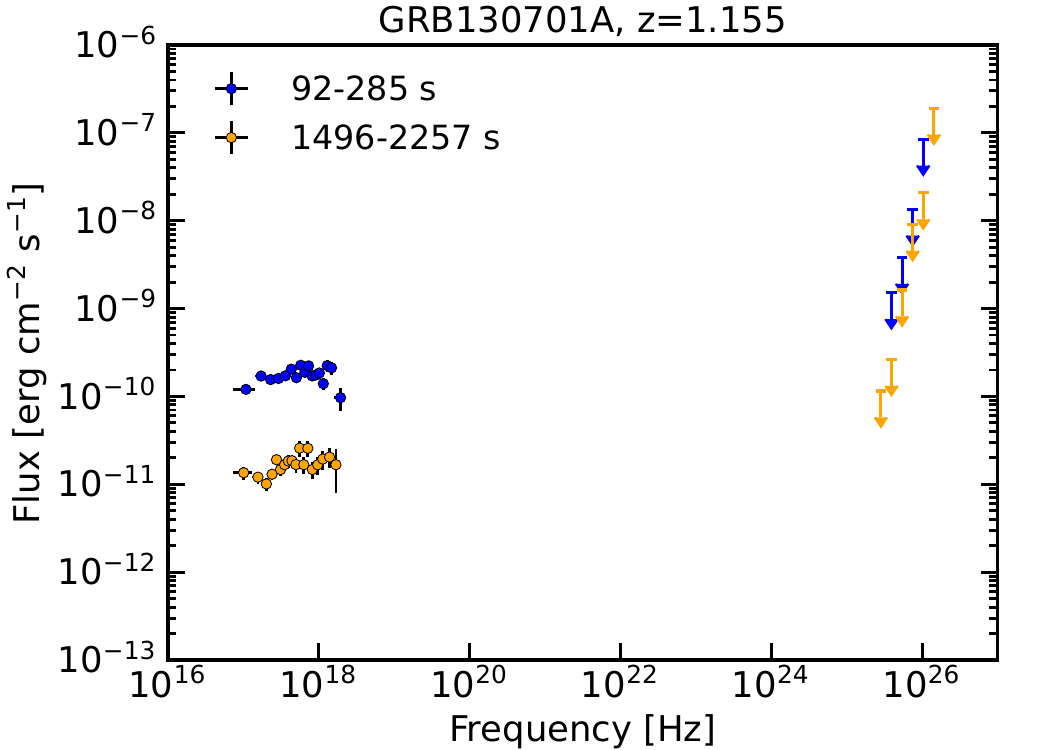} 
\includegraphics[width=0.48\textwidth]{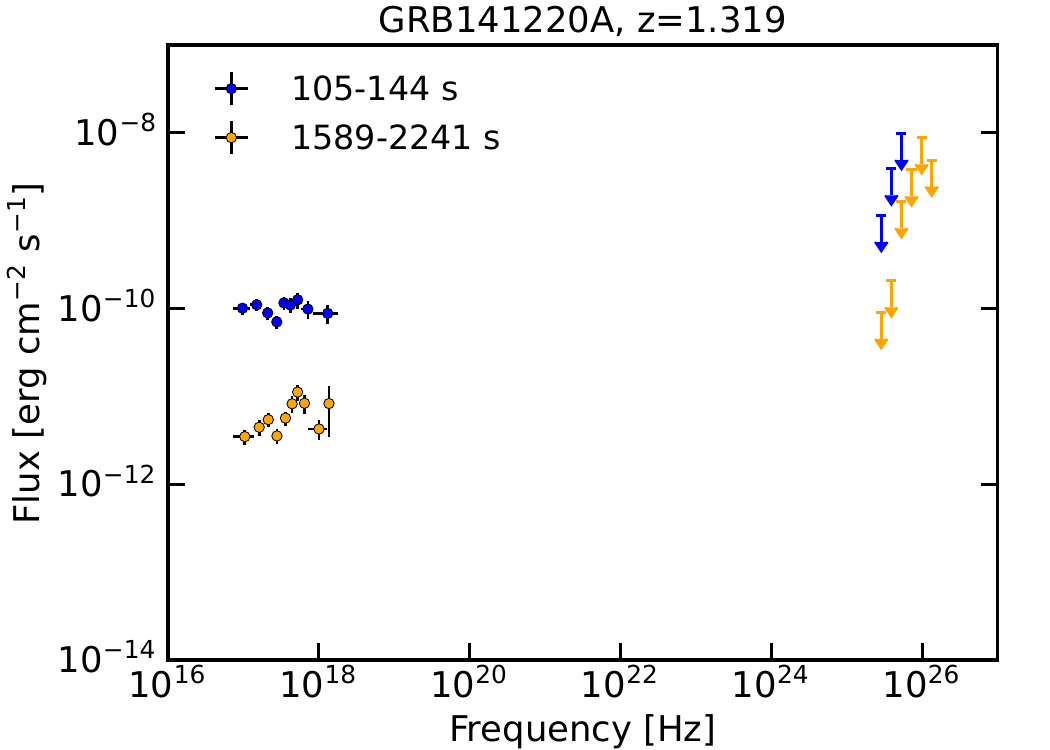}
\caption{Simultaneous X-ray and MAGIC SEDs for GRB~130701A and GRB~141220A. The X-ray fluxes are corrected for dust extinction. The VHE flux ULs are corrected for EBL absorption. For each GRB, two different time intervals are considered. }
\label{fig:interesting_grbs2}
\end{figure*}

These results show a few interesting cases in which the de-absorbed VHE flux UL lies very close to the average XRT flux or even below it, depending on the assumptions. Since this comparison is based on integrated fluxes, we chose the most promising cases (GRB130701A and GRB141220A) and performed a spectral analysis. XRT spectra and MAGIC differential ULs were extracted over the time ranges that included observations in both bands. XRT spectra were extracted with the online tool available at the online repository \footnote{https://www.swift.ac.uk/xrt\_curves/} and were analysed with the XSPEC software. We modelled each analysed XRT spectrum with an absorbed power law accounting both for Galactic and intrinsic metal absorption using the XSPEC models \textit{tbabs} and \textit{ztbabs}, respectively. The Galactic contribution was fixed to the value reported in the automatic analysis tool, and the column density in the host galaxy was a free parameter. The spectral data, rebinned for plotting purposes and de-absorbed for both Galactic and intrinsic absorption, are shown in Fig.~\ref{fig:interesting_grbs2}.

In conclusion, the comparison between MAGIC ULs with simultaneous XRT fluxes shows that VHE flux in many cases cannot be constrained to the same level as the X-ray flux. The comparison of energy-integrated flux light curves seemed to suggest a few cases in which MAGIC ULs can constrain the VHE flux to be lower than the X-ray flux. Nevertheless, a better comparison of the SEDs leaves open the possibility of a VHE emission at the same level or even brighter than the X-ray emission. This indicates that ULs by MAGIC cannot exclude a VHE emission similar to the emission that was detected so far in a handful of GRBs for any of the GRBs presented in this paper, that is, a VHE emission with a luminosity that is comparable to that of the simultaneous X-ray luminosity.

\section{Conclusion}
\label{sec:conclusion}
We presented the main results of the GRB follow-up campaign carried out by MAGIC from 2013 to 2019, which is the seven-year period that starts with an upgrade of the procedures for GRB searches and ends with the year in which the first GRBs at TeV energies were detected. During this period, 66 GRBs were followed-up by MAGIC, and the data taking in 42 cases was not affected by technical or observational problems. The best candidates for detecting a possible VHE component are those observed by MAGIC within the time window defined by $T_{90}$ or, in general, with a short delay ($T_{delay} < 10^4$ s). However, most of the events that fulfilled this criterion have an unknown redshift or a redshift value $z > 1$ (see Fig.~\ref{fig:delay_t90_ratio}), which prevented us from inferring meaningful constraints on the presence of a VHE component. The only two cases of early observations of GRBs with $z < 1$ are GRB\,190114C (detected by MAGIC) and GRB\,160821B (showing a possible hint of a detection). 

We focused on the MAGIC data analysis of the 39 non-detected GRBs (excluding 3 events with a detection or a hint of a detection) that were followed-up from 2013 to 2019. We estimated the flux ULs at the 95\% confidence level following a different approach depending on the available information for the event. For the bursts with an unknown redshift, with a redshift $z \geq 2$, or  events that were observed at zenith angle Zd $ > 40$ deg, which are 33 out of 39 GRBs , we computed the night-wise observed flux ULs and we compared these values with the published results of the GRBs detected in the VHE domain and the $2\sigma$ level sensitivities of the MAGIC and CTAO-North array at two reference energy values, 150 GeV and 250 GeV. The comparison with the detected GRBs did not reveal any particular difference in terms of their intrinsic properties. The observed ULs are well below the flux points derived from the detected GRBs, and they lie at the level of the $2\sigma$ sensitivity of MAGIC and above the CTAO-North one. This indicates that the non-detected GRBs are simply fainter or at larger distances and that they are thus more affected by EBL absorption than the detected GRBs.

For the sample of GRBs with a redshift $z < 2$ and that were observed at a zenith angle Zd $< 40$ deg , which are 6 out of 39 GRBs, we computed EBL-corrected flux ULs in selected energy and time intervals and for different assumptions on the intrinsic gamma-ray spectrum (a power law with photon indices 1.6 and 2.2) and EBL absorption models (D11, F18, and G12). We compared the MAGIC de-absorbed flux ULs with the flux in the soft X-ray band. This comparison provided relevant information to address the open questions of the possible universality of the GRB VHE emission component and the connection with the other lower-energy bands, especially with the X-ray band. We plotted the de-absorbed XRT flux light curves, the XRT average flux in the MAGIC observational time window, and the derived MAGIC flux ULs. For a few cases, we also added the available information on the LAT estimated flux in the 0.1 - 100 GeV band. This comparison showed that ULs are above the corresponding flux in the X-ray band. Only two cases (GRB130701A and GRB141220A) indicated that the VHE flux might be at the same level or below the average XRT flux. For these cases, we performed a refined spectral analysis of the X-ray and VHE simultaneous data and displayed the SEDs in both bands. This further comparison showed that the VHE ULs do not constrain the TeV component at the same level as or below the X-ray one.

In conclusion, these results confirm that MAGIC flux ULs cannot exclude the presence of a VHE emission component with properties similar to those detected in the current population of TeV GRBs. In particular, the possibility that the luminosity in the VHE domain is similar to that in the X-ray band is still open. For the most constraining cases of this study, we found that the VHE component in the MAGIC energy range was constrained to be no more than five to ten times brighter than the simultaneous X-ray emission. Fig.~\ref{fig:non_interesting_grbs} showed that the expected improved sensitivity of CTAO will allow us to fill this gap and lead to an increased number of detections, as well as to more constraining ULs providing crucially important information on a larger fraction of the GRB population. 

\begin{acknowledgements}
We would like to thank the Instituto de Astrof\'{\i}sica de Canarias for the excellent working conditions at the Observatorio del Roque de los Muchachos in La Palma. The financial support of the German BMBF, MPG and HGF; the Italian INFN and INAF; the Swiss National Fund SNF; the grants PID2019-107988GB-C22, PID2022-136828NB-C41, PID2022-137810NB-C22, PID2022-138172NB-C41, PID2022-138172NB-C42, PID2022-138172NB-C43, PID2022-139117NB-C41, PID2022-139117NB-C42, PID2022-139117NB-C43, PID2022-139117NB-C44, CNS2023-144504 funded by the Spanish MCIN/AEI/ 10.13039/501100011033 and "ERDF A way of making Europe; the Indian Department of Atomic Energy; the Japanese ICRR, the University of Tokyo, JSPS, and MEXT; the Bulgarian Ministry of Education and Science, National RI Roadmap Project DO1-400/18.12.2020 and the Academy of Finland grant nr. 320045 is gratefully acknowledged. This work was also been supported by Centros de Excelencia ``Severo Ochoa'' y Unidades ``Mar\'{\i}a de Maeztu'' program of the Spanish MCIN/AEI/ 10.13039/501100011033 (CEX2019-000920-S, CEX2019-000918-M, CEX2021-001131-S) and by the CERCA institution and grants 2021SGR00426 and 2021SGR00773 of the Generalitat de Catalunya; by the Croatian Science Foundation (HrZZ) Project IP-2022-10-4595 and the University of Rijeka Project uniri-prirod-18-48; by the Deutsche Forschungsgemeinschaft (SFB1491) and by the Lamarr-Institute for Machine Learning and Artificial Intelligence; by the Polish Ministry Of Education and Science grant No. 2021/WK/08; and by the Brazilian MCTIC, CNPq and FAPERJ.
LN acknowledges funding by the European Union-Next Generation EU, PRIN 2022 RFF M4C21.1 (202298J7KT - PEACE)\\
\textit{Author contribution}: A. Berti: MAGIC data analysis, UL calculation, statistics on the triggers and AAS maintenance, paper editing; F. Longo: coordination of the project, discussion on the interpretation, paper editing; D. Miceli: coordination of results, MAGIC data analysis, UL derivation, multi-wavelength interpretation, leading paper drafting, writing and editing; L. Nava: coordination and leading multi-wavelength interpretation, X-ray data analysis, paper editing and drafting. A. Carosi, E. Moretti, K. Noda, A. Donini, A. Fattorini, S. Fukami, J. Green, Y. Suda: MAGIC data analysis; The rest of the authors have contributed in one or several of the following ways: design, construction, maintenance and operation of the instrument(s); preparation and/or evaluation of the observation proposals; data acquisition, processing, calibration and/or reduction; production of analysis tools and/or related Monte Carlo simulations; discussion and approval of the contents of the draft.

\end{acknowledgements}

\bibliographystyle{aa} 
\bibliography{aa55468-25.bib}

\begin{appendix}

\onecolumn

\section{GRBs observed by MAGIC between 2013 and 2019}
\begin{table*}[!h]
\begin{center}
\caption{\centering List of GRBs observed by MAGIC (under acceptable conditions) between 2013 and 2019.}
\begin{tabular}{cccccccc}
\toprule
GRB & Redshift&  Instrument & $T_{\rm 90}$ & $T_0$  & $T_{\rm start}$ & $T_{\rm delay}$ & Zenith angle  \\
name & &(position) & \small{[s]}  & \small{[UTC]} & \small{[UTC]} & \small{[s]} & \small{[deg]}\\
\midrule
130502A & &\textit{Swift}-BAT  &  3  & 17:50:30  & 20:57:03       &  11193  & 33.9-40.1       \\
130504A & &\textit{Swift}-BAT  & 50  & 02:05:34  & 02:13:09       &  455    & 44.7-56.5       \\
130606A & 5.913&\textit{Swift}-BAT  & 277 & 21:04:39  & 21:15:28       &  649    & 1.7-46.1        \\
130612A & 2.006&\textit{Swift}-BAT  & 5.6 & 03:22:22  & 03:23:08       &  46     & 38.0-53.0       \\
130701A & 1.155&\textit{Swift}-BAT  & 4.4 & 04:17:43  & 04:18:32       &  49     & 15.9-22.6       \\
130903A & &INTEGRAL            & 69 & 00:47:20  & 03:57:32       &  11412  & 51.9-62.8       \\
131030A & 1.295&\textit{Swift}-BAT  & 41 & 20:56:19  & 20:56:45       &  26     & 33.7-39.7       \\
140430A & 1.60&\textit{Swift}-BAT  & 174 & 20:33:36  & 20:52:06       &  1110   & 45.6-73.3       \\
140709A & &\textit{Swift}-BAT  & 98.6 & 01:13:41  & 03:22:13       &  7712   & 24.6-37.0       \\
140930B & &\textit{Swift}-BAT  & 0.84 & 19:41:42  & 21:10:05       &  5303   & 18.8-51.4       \\
141026A & 3.35&\textit{Swift}-BAT  & 146 & 02:36:51  & 02:38:27       &  96     & 16.3-54.1      \\
141220A & 1.32&\textit{Swift}-BAT  & 7.21 & 06:02:52  & 06:03:47       &  55     & 18.9-24.0      \\
150213A & &\textit{Fermi}-GBM  & 4.1 & 00:01:48  & 00:03:08       &  80     & 48.2-60.6     \\
150428A & &\textit{Swift}-BAT  & 53.2 & 01:30:40  & 01:32:11       &  91     & 27.0-57.7      \\
150428B & &\textit{Swift}-BAT  & 131 & 03:12:03  & 03:13:03       &  60     & 27.0-57.7       \\
150819A & &\textit{Swift}-BAT  & 52.1 & 00:50:08  & 02:11:51       &  4903   & 37.4-54.4       \\
151118A & &\textit{Swift}-BAT  & 23.4 & 03:06:30  & 03:07:14       &  44     & 42.8-57.4      \\
151215A & 2.59&\textit{Swift}-BAT  & 17.8 & 03:01:28  & 03:01:58       &  30     & 15.8-58.0       \\
160119A &  &\textit{Swift}-BAT & 116 & 03:06:07  & 03:17:09       &  662    & 13.2-58.7        \\
160310A &  &\textit{Fermi}-LAT & 18.2 & 00:22:57  & 20:30:16       &  72439  & 35.5-40.9          \\
160313A &  &\textit{Swift}-BAT & 42.6 & 02:37:14  & 02:39:01       &  107    & 30.3-53.3         \\
160504A &  &\textit{Swift}-BAT & 53.9 & 19:30:36  & 20:56:29       &  5153   & 26.9-33.7         \\
160509A &  1.17&\textit{Fermi}-LAT & 370 & 08:59:04  & 21:21:07 (+2d) &  217323 & 49.2-72.2          \\
160623A & 0.367& \textit{Fermi}-LAT & 50 & 05:00:34  & 02:05:31       &  75897  & 27.0-54.7          \\
160625B &  1.406&\textit{Fermi}-LAT & 460 & 22:43:24  & 23:29:38       &  2774   & 21.8-54.9         \\
160821B &  0.16&\textit{Swift}-BAT & 0.48 & 22:29:13 & 22:29:37 & 24 & 33.4-43.6 \\
160910A &  &\textit{Fermi}-GBM & 24.3 & 17:19:38  & 20:21:54       &  10936  & 45.4-72.9        \\
160927A &  &\textit{Swift}-BAT & 0.48 & 18:04:49  & 20:03:00       &  7091   & 32.0-58.8        \\
161229A &  &\textit{Fermi}-GBM & 33.5 & 21:03:48  & 23:05:54       &  7326   & 22.0-26.1         \\
170728B &  &\textit{Swift}-BAT & 47.7 & 23:03:19  & 23:03:58       &  39     & 41.8-52.7        \\
170921B &  &\textit{Fermi}-GBM & 39.4 & 04:02:11  & 04:48:04       &  2753   & 48.4-60.6         \\
171020A & 1.87  &\textit{Swift}-BAT & 41.9 & 23:07:09  & 23:08:37       &  88     & 13.5-34.9        \\
171210A &  &\textit{Fermi}-LAT & 12 & 11:49:15  & 20:33:11       &  31436  & 30.9-61.9        \\
180512A &  &\textit{Swift}-BAT & 24.0 & 22:01:46  & 22:03:11       &  85    & 7.6-38.4         \\
180715A &  &\textit{Swift}-BAT & 0.68 & 18:07:05  & 21:27:24       &  12019  & 27.9-34.5         \\
180720C &  &\textit{Swift}-BAT & 124.2 & 22:23:57  & 22:25:44       &  107    & 55.3-55.4         \\
180904A &  &\textit{Swift}-BAT & 5.39 & 21:28:32  & 21:30:07       &  95     & 23.7-60.2        \\
181225A &  &\textit{Fermi}-LAT & 41.5 & 11:44:10  & 19:56:06 (+1d) &  115916 & 46.7-62.6      \\
190106B &  &\textit{Fermi}-GBM & 11.8 & 20:47:10  & 20:49:13       &  123    & 60.0-60.4          \\
190114C &  0.425&\textit{Swift}-BAT & 25$^*$ & 20:57:03  & 20:58:01       &  58     & 55.6-80.0       \\
190829A &  0.078&\textit{Swift}-BAT  & 62.9 & 19:55:53 & 02:23:48 (+2d) & 109624 & 37.7-59.6  \\
191004A &  &\textit{Swift}-BAT   & 2.44  & 18:07:02 & 00:42:30  &  23728  & 65.4-69.9    \\
\bottomrule
\end{tabular}
\end{center}
\tablefoot{The GRB redshift, when measured, is reported in the second column.
The third column reports the name of the satellite which provided the sky coordinates (e.g.\ through GCN Notices or Circulars). The forth column reports the prompt emission duration $T_{90}$. The fifth and sixth columns give the time of the trigger $T_0$ and the time $T_{\rm start}$ when MAGIC started the observations. The delay $T_{\rm delay}$(in seconds) is computed as the difference between the start time and the trigger time. The last column reports the zenith angle range related to each GRB observation. $^*$Nominally, the $T_{90}$ of GRB 190114C measured by and GBM is $>300$\,s. However, most of the emission recorded by and GBM was interpreted as afterglow radiation. The end of the prompt emission can be roughly identified with the end of flux variability, which occurs approximately at 25\,s.} 
\label{tab:grb} 
\end{table*}

\section{Observed flux UL for GRBs with unknown $z$ or $z \geq 2$ or Zd $> 40$ deg}

\begin{longtable}{p{3cm}ccc|cc}
\caption{\centering Night-wise flux ULs on the observed flux for the subsample of GRBs with unknown $z$ or $z \geq 2$ or Zd $ > 40$ deg. }\\
\hline
\text{GRB} & \text{{T$_{\text{obs}}$}} & \text{{T$_{\text{start}}$ - T$_{\text{stop}}$}} & \text{$E$} & \textbf{$\alpha = 3.5$} & \textbf{$\alpha = 5.5$} \\ 
\textbf{} & & & \textbf{} & \tiny{$10^{-12}$} & \tiny{$10^{-12}$ } \\
\text{name} & \text{[s]} & \text{[s]} & \text{[TeV]} & \tiny{[TeV cm$^{-2}$ s$^{-1}$]} & \tiny{[TeV cm$^{-2}$ s$^{-1}$]} \\
\hline
\endfirsthead
\caption{continued}\\
\hline
\text{GRB} & \text{{T$_{\text{obs}}$}} & \text{{T$_{\text{start}}$ - T$_{\text{stop}}$}} & \text{$E$} & \textbf{$\alpha = 3.5$} & \textbf{$\alpha = 5.5$} \\ 
\textbf{} & & & \textbf{} & \tiny{$10^{-12}$} & \tiny{$10^{-12}$ } \\
\text{name} & \text{[s]} & \text{[s]} & \text{[TeV]} & \tiny{[TeV cm$^{-2}$ s$^{-1}$]} & \tiny{[TeV cm$^{-2}$ s$^{-1}$]} \\
\hline
\endhead
\hline
\endfoot
\endlastfoot

GRB130502A & 1775 & 11769 - 13599 &   0.16 - 0.22 & 14.6  & 10.9 \\ 
& & & 0.22 - 0.30 & 15.1 & 10.2 \\ 
& & &0.30 - 0.41 & 7.41 & 5.36 \\ 
& & &0.41 - 0.55 & 9.81 & 5.15 \\ 
& & &0.55 - 0.75 & 8.01 & 4.38 \\ 
& & & 0.75 - 1.02 & 21.0 & 9.20 \\ 
\hline
GRB130504A & 10481 & 455 - 11487 & 0.22 - 0.30 & 4.31 & 2.95 \\ 
& & &0.30 - 0.41 & 4.06 & 2.89 \\ 
& & &0.41 - 0.55 & 2.48 & 1.64 \\ 
& & &0.55 - 0.75 & 2.02 & 1.07 \\ 
& & &0.75 - 1.02 & 1.86 & 1.02 \\ 
\hline
GRB130606A & 10704 & 1747 - 12868 &  0.12 - 0.16 & 5.26 & 3.88 \\ 
& & & 0.16 - 0.22 & 4.24 & 3.23 \\ 
& & & 0.22 - 0.30 & 5.66 & 4.40 \\ 
& & & 0.30 - 0.41 & 2.95 & 2.31 \\ 
& & & 0.41 - 0.55 & 2.34 & 1.62 \\ 
& & & 0.55 - 0.75 & 1.74 & 0.93 \\ 
& & & 0.75 - 1.02 & 4.82 & 2.54 \\ 
\hline
GRB130612A & 3822 & 688 - 4664 & 0.16 - 0.22 & 15.8 & 11.4 \\ 
& & & 0.22 - 0.30 & 7.16 & 5.27 \\ 
& & & 0.30 - 0.41 & 12.7 & 7.63 \\ 
& & & 0.41 - 0.55 & 5.11 & 2.98 \\ 
& & & 0.55 - 0.75 & 4.20 & 1.76 \\ 
& & & 0.75 - 1.02 & 3.57 & 1.53 \\ 
\hline
GRB130903A & 2560 & 11412 - 14476 & 0.41 - 0.55 & 9.40 & 6.18 \\ 
& & & 0.55 - 0.75 & 11.2 & 6.74 \\ 
& & & 0.75 - 1.02 & 7.50 & 4.10 \\ 
\hline
GRB140430A & 1859 & 1110 - 3018 & 0.22 - 0.30 & 8.91 & 6.02 \\ 
& & & 0.30 - 0.41 & 7.62 & 5.05 \\ 
& & & 0.41 - 0.55 & 5.85 & 3.68 \\ 
& & & 0.55 - 0.75 & 4.93 & 2.20 \\ 
& & & 0.75 - 1.02 & 10.1 & 3.68 \\ 
\hline
GRB140709A & 5834 & 7712 - 13745 &  0.12 - 0.16 & 14.5 & 10.7 \\ 
& & & 0.16 - 0.22 & 14.5 & 10.7 \\ 
& & & 0.22 - 0.30 & 3.98 & 3.16 \\ 
& & & 0.30 - 0.41 & 4.45 & 3.16 \\ 
& & & 0.41 - 0.55 & 2.37 & 1.97 \\ 
& & & 0.55 - 0.75 & 8.15 & 5.21 \\ 
& & & 0.75 - 1.02 & 7.75 & 4.55 \\ 
\hline
GRB140930B & 7839 & 5690 - 14073 & 0.22 - 0.30 & 8.21 & 6.66 \\ 
& & & 0.30 - 0.41 & 4.48 & 3.63 \\ 
& & & 0.41 - 0.55 & 4.05 & 3.20 \\ 
& & & 0.55 - 0.75 & 2.70 & 1.89 \\ 
& & & 0.75 - 1.02 & 2.77 & 1.89 \\ 
\hline
GRB141026A & 6482 & 223 - 10818 & 0.16 - 0.22 & 5.39 & 3.95 \\ 
& & & 0.22 - 0.30 & 4.00 & 3.02 \\ 
& & & 0.30 - 0.41 & 3.57 & 2.50 \\ 
& & & 0.41 - 0.55 & 2.55 & 1.81 \\ 
& & & 0.55 - 0.75 & 1.90 & 1.09 \\ 
& & & 0.75 - 1.02 & 5.41 & 1.93 \\ 
\hline
GRB150213A & 3727 & 80 - 5334 &  0.30 - 0.41 & 8.57 & 6.02 \\ 
& & & 0.41 - 0.55 & 10.9 & 7.67 \\ 
& & & 0.55 - 0.75 & 4.87 & 2.72  \\ 
& & & 0.75 - 1.02 & 6.05 & 2.74 \\ 
\hline
GRB150428A & 3503 & 122 - 3713 & 0.22 - 0.30 & 4.94 & 3.34 \\ 
& & & 0.30 - 0.41 & 2.84 & 2.15 \\ 
& & & 0.41 - 0.55 & 6.67 & 3.67 \\ 
& & & 0.55 - 0.75 & 7.16 & 5.83 \\ 
& & & 0.75 - 1.02 & 8.05 & 5.25 \\ 
\hline
GRB150428B & 7467 & 330 - 8200 & 0.16 - 0.22 & 3.83 & 2.80 \\ 
& & & 0.22 - 0.30 & 3.34 & 2.44 \\ 
& & & 0.30 - 0.41 & 2.40 & 1.77 \\ 
& & & 0.41 - 0.55 & 2.40 & 1.77 \\ 
& & & 0.55 - 0.75 & 4.74 & 3.04 \\ 
& & & 0.75 - 1.02 & 5.06 & 1.26 \\ 
\hline
GRB150819A & 5103 & 4907 - 10211 & 0.22 - 0.30 & 7.99 & 5.46 \\ 
& & & 0.30 - 0.41 & 11.2 & 7.65 \\ 
& & & 0.41 - 0.55 & 4.61 & 2.64 \\ 
& & & 0.55 - 0.75 & 4.99 & 2.84 \\ 
& & & 0.75 - 1.02 & 7.56 & 3.17 \\ 
\hline
GRB151118A & 10689 & 44 - 11216 & 0.22 - 0.30 & 9.13 & 6.23 \\ 
& & & 0.30 - 0.41 & 3.74 & 2.55 \\ 
& & & 0.41 - 0.55 & 2.82 & 1.61 \\ 
& & & 0.55 - 0.75 & 2.40 & 1.41 \\ 
& & & 0.75 - 1.02 & 2.14 & 0.88 \\ 
\hline
GRB151215A & 10403 & 30 - 12434 &  0.12 - 0.16 & 8.52 & 6.07 \\ 
& & & 0.16 - 0.22 & 4.79 & 3.54 \\ 
& & & 0.22 - 0.30 & 7.40 & 5.43 \\ 
& & & 0.30 - 0.41 & 5.52 & 3.79 \\ 
& & & 0.41 - 0.55 & 2.81 & 1.72 \\ 
& & & 0.55 - 0.75 & 2.29 & 1.29 \\ 
& & & 0.75 - 1.02 & 2.24 & 0.76 \\ 
\hline
GRB160119A & 5858 & 1888 - 7954 &  0.16 - 0.22 & 11.0 & 7.99 \\ 
& & & 0.22 - 0.30 & 7.29 & 5.30 \\ 
& & & 0.30 - 0.41 & 3.84 & 2.42 \\ 
& & & 0.41 - 0.55 & 3.25 & 2.15 \\ 
& & & 0.55 - 0.75 & 3.83 & 2.46 \\ 
& & & 0.75 - 1.02 & 3.26 & 1.79 \\ 
\hline
GRB160310A & 2343 & 72440 - 74847 &  0.12 - 0.16 & 30.2 & 19.8 \\ 
& & & 0.16 - 0.22 & 15.4 & 9.59 \\ 
& & & 0.22 - 0.30 & 8.91 & 6.65 \\ 
& & & 0.30 - 0.41 & 9.05 & 5.76 \\ 
& & & 0.41 - 0.55 & 8.50 & 5.45 \\ 
& & & 0.55 - 0.75 & 4.72 & 2.52 \\ 
& & & 0.75 - 1.02 & 12.4 & 7.72 \\ 
\hline
GRB160310A & 3495 & 159187 - 162841 &  0.12 - 0.16 & 13.6 & 8.62 \\ 
& & & 0.16 - 0.22 & 8.81 & 6.71 \\ 
& & & 0.22 - 0.30 & 6.69 & 5.55 \\ 
& & & 0.30 - 0.41 & 6.74 & 5.14 \\ 
& & & 0.41 - 0.55 & 3.61 & 2.87 \\ 
& & & 0.55 - 0.75 & 3.66 & 2.45 \\ 
& & & 0.75 - 1.02 & 10.0 & 5.27 \\ 
\hline
GRB160310A & 2687 & 332239 - 335312 &  0.16 - 0.22 & 8.69 & 5.74 \\ 
& & & 0.22 - 0.30 & 8.11 & 5.70 \\ 
& & & 0.30 - 0.41 & 6.93 & 4.75 \\ 
& & & 0.41 - 0.55 & 4.51 & 2.84 \\ 
& & & 0.55 - 0.75 & 3.65 & 2.39 \\ 
& & & 0.75 - 1.02 & 9.51 & 5.93 \\ 
\hline
GRB160313A & 11622 & 107 - 12104 & 0.16 - 0.22 & 7.76 & 5.41 \\ 
& & & 0.22 - 0.30 & 5.57 & 3.99 \\ 
& & & 0.30 - 0.41 & 5.57 & 3.99 \\ 
& & & 0.41 - 0.55 & 3.35 & 2.13 \\ 
& & & 0.55 - 0.75 & 2.74 & 1.22 \\ 
& & & 0.75 - 1.02 & 4.22 & 1.29 \\ 
\hline
GRB160504A & 8665 & 5153 - 14126 &  0.12 - 0.16 & 7.60 & 5.76 \\ 
& & & 0.16 - 0.22 & 7.74 & 6.97 \\ 
& & & 0.22 - 0.30 & 7.74 & 6.97 \\ 
& & & 0.30 - 0.41 & 3.42 & 2.75 \\ 
& & & 0.41 - 0.55 & 3.34 & 2.90 \\ 
& & & 0.55 - 0.75 & 2.39 & 1.95 \\ 
& & & 0.75 - 1.02 & 2.19 & 1.84 \\ 
\hline
GRB160509A & 10647 & 315153 - 326820 & 0.30 - 0.41 & 5.74 & 3.28 \\ 
& & & 0.41 - 0.55 & 3.93 & 2.37 \\ 
& & & 0.55 - 0.75 & 2.70 & 1.37 \\ 
& & & 0.75 - 1.02 & 2.28 & 1.12 \\ 
\hline
GRB160509A & 2218 & 407122 - 409400 & 0.30 - 0.41 & 21.7 & 12.9 \\ 
& & & 0.41 - 0.55 & 7.65 & 4.15 \\ 
& & & 0.55 - 0.75 & 9.09 & 3.91 \\ 
& & & 0.75 - 1.02 & 5.29 & 2.24 \\ 
\hline
GRB160927A & 6773 & 7091 - 14145 &  0.16 - 0.22 & 8.63 & 5.12 \\ 
& & & 0.22 - 0.30 & 7.78 & 5.17 \\ 
& & & 0.30 - 0.41 & 5.99 & 4.40 \\ 
& & & 0.41 - 0.55 & 4.15 & 3.21 \\ 
& & & 0.55 - 0.75 & 3.56 & 2.48 \\ 
& & & 0.75 - 1.02 & 2.99 & 1.79 \\ 
\hline
GRB161229A & 4207 & 7326 - 12275 &  0.09 - 0.12 & 8.93 & 6.19 \\ 
& & & 0.12 - 0.16 & 6.74 & 4.94 \\ 
& & & 0.16 - 0.22 & 6.45 & 5.12 \\ 
& & & 0.22 - 0.30 & 3.89 & 2.86 \\ 
& & & 0.30 - 0.41 & 3.53 & 2.82 \\ 
& & & 0.41 - 0.55 & 6.08 & 4.20 \\ 
& & & 0.55 - 0.75 & 3.34 & 2.61 \\ 
& & & 0.75 - 1.02 & 4.32 & 3.37 \\ 
\hline
GRB170728B & 2774 & 39 - 3017 & 0.22 - 0.30 & 6.03 & 3.95 \\ 
& & & 0.30 - 0.41 & 6.82 & 4.48 \\ 
& & & 0.41 - 0.55 & 4.48 & 1.96 \\ 
& & & 0.55 - 0.75 & 3.29 & 1.20 \\ 
& & & 0.75 - 1.02 & 4.58 & 1.38 \\ 
\hline
GRB170728B & 5830 & 80279 - 86356 & 0.16 - 0.22 & 8.16 & 5.57 \\ 
& & & 0.22 - 0.30 & 5.34 & 3.56 \\
& & & 0.30 - 0.41 & 2.82 & 2.01 \\ 
& & & 0.41 - 0.55 & 2.95 & 1.79 \\ 
& & & 0.55 - 0.75 & 3.51 & 2.42 \\ 
& & & 0.75 - 1.02 & 5.32 & 1.99 \\ 
\hline
GRB170921B & 2303 & 2792 - 6853 & 0.30 - 0.41 & 10.7 & 6.64 \\ 
& & & 0.41 - 0.55 & 11.4 & 6.55 \\ 
& & & 0.55 - 0.75 & 5.09 & 2.76 \\ 
& & & 0.75 - 1.02 & 4.46 & 1.86 \\ 
\hline
GRB171210A & 8102 & 31436 - 39959 &  0.16 - 0.22 & 12.8 & 8.35 \\ 
& & & 0.22 - 0.30 & 5.93 & 4.06 \\ 
& & & 0.30 - 0.41 & 4.59 & 2.69 \\ 
& & & 0.41 - 0.55 & 4.47 & 2.67 \\ 
& & & 0.55 - 0.75 & 2.63 & 0.97 \\ 
& & & 0.75 - 1.02 & 4.04 & 1.42 \\ 
\hline
GRB180512A & 13807 & 85 - 14519 &  0.09 - 0.12 & 9.72 & 6.35 \\ 
& & & 0.12 - 0.16 & 7.28 & 5.21 \\ 
& & & 0.16 - 0.22 & 3.34 & 2.57 \\ 
& & & 0.22 - 0.30 & 2.49 & 1.91 \\ 
& & & 0.30 - 0.41 & 2.30 & 1.74 \\ 
& & & 0.41 - 0.55 & 2.23 & 1.61 \\ 
& & & 0.55 - 0.75 & 1.48 & 0.98 \\ 
& & & 0.75 - 1.02 & 3.62 & 2.36 \\ 
\hline
GRB180715A & 3498 & 12019 - 15675 &  0.12 - 0.16 & 8.92 & 6.03 \\ 
& & & 0.16 - 0.22 & 6.58 & 4.99 \\ 
& & & 0.22 - 0.30 & 5.14 & 3.68 \\ 
& & & 0.30 - 0.41 & 4.54 & 2.33 \\ 
& & & 0.41 - 0.55 & 3.34 & 2.15 \\ 
& & & 0.55 - 0.75 & 7.64 & 3.39 \\ 
& & & 0.75 - 1.02 & 8.37 & 5.97 \\ 
\hline
GRB180720C & 2102 & 151 - 2507 &  0.75 - 1.02 & 8.22 & 6.38 \\ 
\hline
GRB180904A & 13834 & 212 - 14758 &  0.12 - 0.16 & 7.54 & 5.04 \\ 
& & & 0.16 - 0.22 & 7.17 & 4.97 \\ 
& & & 0.22 - 0.30 & 4.20 & 2.96 \\ 
& & & 0.30 - 0.41 & 2.95 & 1.92 \\ 
& & & 0.41 - 0.55 & 4.37 & 2.65 \\ 
& & & 0.55 - 0.75 & 2.11 & 1.09 \\ 
& & & 0.75 - 1.02 & 1.77 & 0.49 \\ 
\hline
GRB181225A & 5235 & 115916 - 121415 & 0.30 - 0.41 & 10.8 & 7.44 \\ 
& & & 0.41 - 0.55 & 6.09 & 4.42 \\ 
& & & 0.55 - 0.75 & 4.21 & 3.15 \\ 
& & & 0.75 - 1.02 & 3.65 & 2.78 \\ 
\hline
\label{tab:observed_uls}
\end{longtable}
\tablefoot{Flux ULs are in units of $10^{-12}$\,TeV\,cm$^{-2}$\,s$^{-1}$. The observed photon indices $\alpha$ assumed is mentioned (3.5
and 5.5). In the second and third column of the table, the total observation time $T_{obs}$ and the time interval ($T_{start}$ and $T_{stop}$) in which ULs are estimated with respect to the burst trigger time $T_0$ are reported. Multiple entries for the GRBs refer to different observational nights.}

\end{appendix}

\end{document}